# Highly entangled magnetodielectric, magnetostriction, and spin-phonon coupling phenomena in the $S = 1$ triangular-lattice antiferromagnet $Ni_2ScSbO_6$

Neha Patel[1], Arkadeb Pal[2,3†], C. W. Wang[4], G. R. Blake[3], J. Khatua[5], T. W. Yen[2], Susaiammal Arokiasamy[6], H. S. Kunwar[7], Y. C. Lai[4], Y. C. Chuang[4], V. Sathe[7], Kwang-Yong Choi[5], H. D. Yang[2], Sandip Chatterjee[1*]

[1]*Department of Physics, Indian Institute of Technology (BHU) Varanasi, 221005, India*

[2]*Department of Physics, National Sun Yat-sen University, Kaohsiung 80424, Taiwan*

[3]*Zernike Institute for Advanced Materials, University of Groningen, 9747 AG Groningen, The Netherlands*

[4]*National Synchrotron Radiation Research Center, Hsinchu, 30076, Taiwan*

[5]*Department of Physics, Sungkyunkwan University, Suwon 16419, Republic of Korea*

[6]*Department of Physics, St. Bede's College, Navbahar, Shimla 171002, India*

[7]*UGC-DAE Consortium for Scientific Research, University Campus, Khandwa Road, Indore 452001, India*

Corresponding authors: †pal.arkadeb@gmail.com; *schaterji.app@iitbhu.ac.in

## ABSTRACT

Magnetic compounds with noncentrosymmetric chiral crystal structures and spin-frustrated lattices are renowned for their complex magnetic ordering and diverse correlated phenomena. In this report, we present a comprehensive study of the chiral and triangular lattice magnetic system $Ni_2ScSbO_6$, which exhibits incommensurate spin-density wave antiferromagnetic long-range ordering at a temperature of $T_N = 62$ K, as revealed by bulk magnetization, specific heat, and neutron diffraction studies. In contrast to its isostructural materials $Ni_3TeO_6$ and $Ni_2InSbO_6$, $Ni_2ScSbO_6$ does not exhibit ferroelectric ordering, potentially due to its unique sinusoidal spin-density wave magnetic ordering. The onset of this magnetic ordering is closely linked to a series of intriguing phenomena occurring at $T_N$, including prominent magnetodielectric coupling manifested by a sharp λ-like dielectric peak, substantial spin-phonon coupling characterized by anomalous phonon softening, and a remarkable volume magnetostriction effect revealed by high-resolution synchrotron X-ray diffraction studies. These phenomena demonstrate an intricately coupled behaviour, highlighting $Ni_2ScSbO_6$ as a unique and interesting material.

## INTRODUCTION

The quest for advanced, energy-efficient spintronics devices has sparked a surge in investigations focused on multifunctional materials that exhibit a diverse array of mutually entangled properties. The close correspondence among various microscopic degrees of freedom in these systems - such as spin, orbital, charge, lattice, dipole, and phonon–lays the groundwork for controlling one property through the conjugate field of another [1–4]. In this context, magnetoelectric and magnetodielectric (MD) materials have garnered substantial attention for their potential to realize practical applications, such as tunable filters, magnetic sensors, and spin-charge transducers [5–8]. While both intrinsic and extrinsic mechanisms (such as the Maxwell–Wagner effect) have been proposed to explain the origin of the MD effect, intrinsic mechanisms hold promise for envisioning devices with faster switching and reduced losses. Apart from such technological prospects, investigating MD properties across various classes of magnets also contributes to our understanding of coupled properties that stem from the inherent antagonism between spin and dipolar origins. The exploration of systems with complex magnetic ordering can be an efficacious means towards crafting novel MD and/or magnetoelectric materials.

On the other hand, fascinating effects emerge from the excitations of collective modes in condensed matter, such as spin waves, and phonons coupled with spin ordering, which constitute a fundamental facet of lattices that contain magnetic ions. These phenomena, predominantly governed by magnetic exchange interactions, occur when such materials are exposed to electromagnetic waves [9]. To address the modern need for device miniaturization, such properties offer a promising path for integrated low-power computation and signal processing, owing to their lack of Joule loss. Moreover, the strong entanglement of spin-lattice-phonon degrees of freedom may trigger further intriguing phenomena such as superconductivity, the phonon Hall effect, spin-Seebeck effect, spin-Peierls transitions, magnetoelectric or MD coupling, and the giant thermal Hall effect [10–16].

Insulating systems with a noncentrosymmetric crystal structure and complex spin ordering can host intriguing effects, including multiferroicity, solitons, skyrmions, the linear electro-optic effect, and the piezoelectric effect [17–24]. These magnetic systems have attracted significant attention in the emerging fields of terahertz and sub-terahertz magnonics and antiferromagnetic spintronics. In this context, the AFM magnet $Ni_2ScSbO_6$, part of the chiral corundum structured family related to $Ni_3TeO_6$, is of particular interest. Recent research has focused on compounds derived from $Ni_3TeO_6$, such as $Ni_2InSbO_6$, $Mn_2MnWO_6$, $Mn_2InSbO_6$, and $Co_2InSbO_6$ [25–29]. Despite their similar crystal structures and symmetry, these compounds exhibit distinct spin orderings and varied physical properties. Ivanov *et al.* first synthesized and investigated the compounds $Ni_2ScSbO_6$ and $Ni_2InSbO_6$ for their crystal and spin structures in 2013. Unlike the commensurate collinear AFM order observed in $Ni_3TeO_6$, both $Ni_2ScSbO_6$ and $Ni_2InSbO_6$ were reported to exhibit antiferromagnetically coupled incommensurate helices [30]. Subsequently, $Ni_2InSbO_6$ has been studied extensively by various groups, particularly for its intriguing magnetic and multiferroic properties [26]. On the contrary, despite having similar crystal structure, research on $Ni_2ScSbO_6$ has remained limited, with many of its properties largely unexplored. In this report, we present a comprehensive study of the crystal and spin structures, magnetization, dielectric, magnetodielectric, and lattice dynamics properties of $Ni_2ScSbO_6$.

## II. EXPERIMENTAL METHODS

A polycrystalline sample of $Ni_2ScSbO_6$ was prepared using the conventional solid-state reaction route. For the reactions, dried high-purity oxide powders of NiO, $Sc_2O_3$, and $Sb_2O_5$, with purities exceeding 99.99% (Alfa-Aesar), were mixed as precursors in the appropriate stoichiometric ratio. The mixture was intimately ground for several hours and then formed into pellets. The resulting pellets were initially sintered at 700°C for 24 hours, then ground into fine powder, pressed, and annealed multiple times at temperature intervals of 100°C up to 1200°C. The series of grinding and sintering was repeated until the X-ray diffraction (XRD) pattern exhibited no impurity peaks. Finally, the samples were sintered at 1300°C for 48 hours. The resulting product was greenish yellow in color, which was assessed for phase purity by performing a synchrotron X-ray diffraction (SXRD) measurement at $T$ = 300 K at the Taiwan Photon Source (TPS) 19A beamline of the National Synchrotron Radiation Research Center (NSRRC), Taiwan, using X-rays with a wavelength ($\lambda_i$) of 0.77489 Å. Additionally, $T$-dependent SXRD patterns were collected over the range of 300–20 K. Two setups with different wavelengths were used in two distinct temperature regimes: $\lambda_i$ = 0.77489 Å for $T$ = 300–100 K and $\lambda_i$ = 0.61992 Å for $T$ = 100–20 K. Magnetization measurements were conducted using a commercial SQUID-based magnetometer (MPMS, Quantum Design). Various dielectric measurements were conducted utilizing a precision Agilent E4980A LCR meter coupled with a custom dielectric probe within a Quantum Design MPMS, facilitating control over both magnetic field ($H$) and temperature ($T$). Specific heat measurements ($C_P$) were carried out utilizing a heat-pulsed thermal relaxation calorimeter within a Quantum Design physical property measurement system (PPMS). Raman data were collected using a Horiba Jobin Yvon HR-800 spectrometer equipped with an 1800 g/mm grating, an edge filter for Rayleigh line rejection, and a CCD detector. A 473 nm diode laser served as the excitation source in a backscattering geometry for the Raman measurements. Thermal control was maintained with a Janis ST-500 continuous helium flow cryostat, ensuring a thermal stability of ±0.1 K.

## III. RESULTS AND DISCUSSIONS

### A. Crystal structure

The high-resolution SXRD pattern of $Ni_2ScSbO_6$ acquired at $T$ = 300 K, along with its fitted Rietveld profile, is illustrated in Fig. 1(a). The refinement indicates a single-phase trigonal structure with chiral symmetry $R3$, consistent with prior studies. No impurities were discerned. Detailed crystallographic data, including lattice parameters and Wyckoff positions, are provided in Table SI of the supplemental material (SM) [31], which are congruent with previously reported values [30]. The crystal structure of $Ni_2ScSbO_6$ is illustrated in Figs. 1(b)–(d), depicting interconnected networks of $NiO_6$, $ScO_6$, and $SbO_6$ octahedra. Ni atoms, occupying distinct crystallographic sites Ni(1) and Ni(3), are situated on consecutive layers within the *ab* plane, separated by a distance of 0.1667 *c*. Additionally, Sc and Sb atoms reside on the same planes as Ni(3) and Ni(1), respectively. Thus, the structure comprises two-dimensional (2D) honeycomb layers on the *ab* planes, composed of alternately spaced edge-shared Ni(1)$O_6$ (or Ni(3)$O_6$) and $SbO_6$ (or $ScO_6$) octahedra, as shown in Fig. 1(c). Notably, when examining the nearest neighbor magnetic atoms on the *ab* planes, both Ni(1) and Ni(3) form triangular networks within their respective *ab* planes, as depicted in Fig. 1(c). This triangular lattice arrangement of Ni atoms becomes particularly evident when observing a specific *ab* plane, as demonstrated in Fig. 1(d).

### B. Magnetization

The temperature dependence of the DC magnetization *(M)* measured using zero-field cooled (ZFC) and field cooled (FC) protocols under an applied field of $H$ = 0.05 T is presented in Fig. 2(a). The *M(T)* curves exhibit a sharp λ-like transition at $T_N$ = 62 K, which indicates long-range AFM ordering. The observed $T_N$ is close to, but slightly higher than, the previously reported value (~ 60 K) for this system [30]. At temperatures below 30 K, $M$ slightly increases as $T$ decreases, likely due to the presence of defects or orphan spins that behave like paramagnetic spins, a phenomenon commonly observed in other AFM systems [32,33]. The inverse susceptibility ($1/\chi$) versus $T$ data for $H$ = 2 T was fitted in the paramagnetic region ($T$ = 120 K – 300 K) using the modified Curie-Weiss (CW) law, expressed as: $\chi = \chi_0 + \frac{C}{(T - \theta_{CW})}$ [34]. Here, $\chi_0$ is the $T$–independent susceptibility incorporating two distinct terms: the core diamagnetism arising from the core-electron shells ($\chi_{\text{core}}$) of all the constituent atoms and the open-shell Van Vleck paramagnetism ($\chi_{VV}$) associated with $Ni^{2+}$. $C$ represents the Curie constant, and $\theta_{CW}$ refers to the characteristic CW temperature. The fit is displayed in the inset of Fig. 2(a) and yielded $\chi_0$ = – 9.563×$10^{-4}$ emu/(mole-Oe), an effective paramagnetic moment ($\mu_{eff}$) of 2.94 $\mu_B$/$Ni^{2+}$, and $\theta_{CW}$ = – 87 K. The large negative value of $\theta_{CW}$ indicates that AFM interactions are dominant. The experimentally obtained value of $\mu_{eff}$ closely matches the theoretical spin-only moment ($\mu_{theor}$) of 2.82 $\mu_B$ expected for $Ni^{2+}$ ions. This is in agreement with X-ray absorption spectroscopy (XAS) data at the Ni $L_{2,3}$ edge (see Fig. S2 of SM). Furthermore, an isothermal *M(H)* curve was recorded at $T$ = 2 K, as shown in Fig. 2(b). The linear, non-saturating, and non-hysteretic behavior of the curve clearly suggests that the system possesses overall AFM character.

### C. Specific heat

The $T$-dependence of the zero-field specific heat ($C_p$) is shown in Fig. 2(c), which exhibits a λ-like anomaly at $T_N$ = 62 K, thus suggesting the development of a long-range ordered (LRO) magnetic state. This observation aligns with the magnetization and neutron powder diffraction (NPD) studies (discussed below) on this system. In insulating magnets, the $C_p$ predominantly originates from phonon excitations at high temperatures, while at low temperatures, it is mainly influenced by magnetic interactions. Thus, to determine the magnetic specific heat ($C_{mag}$), the phononic component ($C_{lattice}$) was subtracted from the overall $C_p$. $C_{lattice}$ was estimated by fitting our $C_p(T)$ data in the high-$T$ range ($T$ = 80 –200 K) and then extrapolating down to 2 K, as depicted by the red line in Fig. 2(c). This fitting was performed using the Debye model (with two Debye functions) [35], expressed as : $C_{lattice}(T) = 9k_B\left[\sum_{n=1}^{2} C_n \left(\frac{T}{\theta_{D_n}}\right)^3 \int_0^{\theta_{D_n}/T} \frac{x^4 e^x}{(e^x-1)^2}dx\right]$, where $k_B$ refers to the Boltzmann constant, $\theta_{D_n}$ ($n$ = 1, 2) denotes the characteristic Debye temperatures, and $C_n$ represents the integer coefficients. The best fit, which satisfactorily reproduces the experimentally recorded $C_p(T)$ data at high temperatures, yielded $\theta_{D_1}$ = 290 K and $\theta_{D_2}$ = 920 K. The coefficients $C_1$ and $C_2$ were fixed at a ratio of 1:1.5, representing the ratio of the total number of heavier metal atoms to the total number of lighter oxygen atoms. The resulting $C_{mag}/T$ curve as a function of $T$ is shown in Fig. 2(d). Besides the sharp peak at $T_N$, a broad feature is observed near $T \sim 30$ K, possibly associated with the reentrant glassy state indicated by the *ac*-susceptibility study (Fig. S1 of SM), which may originate from spin frustration and minor lattice defects or disorders. Similar, broad shoulder-like features have previously been reported for other spin-frustrated systems [36]. The magnetic entropy ($S_{mag}$) was estimated by integrating the $C_{mag}/T$ curve, as shown in Fig. 2(d). Approximately 78% of the total $S_{mag}$ is released at $T_N$, and the remainder is recovered above $T_N$. Above $T$ = 80 K, $S_{mag}$ reaches a saturation value of ~9.08 J/mol-K, which matches well with the theoretically expected magnetic entropy ~ 9.13 J/mol-K for $Ni^{2+}$ spins ($S$ = 1). This observation is consistent with the $1/\chi$ *vs.* $T$ curve, which demonstrates a deviation from Curie-Weiss behavior below $T \approx 80$ K (inset of Fig. 2(a)), typical for the presence of short-range magnetic ordering above $T_N$ [34,37]. $C_p(T)$ data were also collected under various applied $H$ up to 7 T, as shown in the inset of Fig. 2(c). The peak associated with $T_N$ is substantially suppressed and shifted towards lower temperature with increasing $H$, which are typical characteristics of AFM systems.

### D. Neutron Powder Diffraction

NPD was performed on $Ni_2ScSbO_6$ under zero-field and applied magnetic fields up to 4 T to gain insight into its spin ordering. As the temperature drops below 60 K, magnetic Bragg peaks appear, with some existing nuclear peaks gaining intensity and two new magnetic reflections emerging. This is evident in the NPD patterns at $T$ = 2 K and 50 K (Fig. 3(a)), where a new pair of magnetic peaks forms on either side of the nuclear peak at 32°. This appears as broadening compared to the nuclear reflections observed at a temperature higher than $T_N$, i.e., 70 K, shown in Fig. 3 (a). To isolate the signal associated with magnetic ordering, the 70 K NPD data were subtracted from the 2 K and 50 K data, as depicted in Fig. 3 (b). The newly emerged magnetic peaks could not be indexed to the fundamental unit cell of the nuclear structure, indicating the presence of an incommensurate magnetically modulated spin structure. The NPD data collected at $T$ = 2 K were analyzed using several candidate incommensurate magnetic models, including amplitude-modulated and orientation-modulated spin structures. Four models exhibited reasonably good fits, as summarized in Table I. The corresponding fitted data are shown in Fig. 3(c) for model 1 and in Fig. S4 [(a) to (c)] of the SM for models 2 to 4, respectively. The spin modulation, characterized by a propagation vector $k$ = (0, 0.03575, 0) relative to the trigonal nuclear unit cell, was obtained from these fittings. As evident from Table I, models 1 to 3 yielded similar goodness-of-fit values, with the best fit achieved for a sinusoidal spin-density wave (SDW). The average refined moment for the SDW model fit is 2.7261×2/π = 1.735 $\mu_B$/$Ni^{2+}$ at 2 K, which is close to the theoretically expected moment of 2 $\mu_B$ for $S$ = 1. The slightly reduced moment value can be attributed to strong quantum fluctuations that prevent the spins from achieving full order, a phenomenon also reported in other systems [38]. In contrast, model 4 exhibited considerably poorer fitting parameters compared to the other three models; thus, it can be ruled out as a possible spin structure for $Ni_2ScSbO_6$. A relatively close values of the goodness-of-fit parameters for the models 1 to 3 indicate that we cannot definitively assert the SDW model as the only viable option, despite its superior fitting parameters. Alternate models 2 and 3, including the helical spin ordering (i.e., model 2) proposed in a previous report [39], may also represent feasible spin configurations in $Ni_2ScSbO_6$. In fact, a similar helical spin structure with slightly different $k$ vectors: $k$ = [0, 0.036(1), 0] for $Ni_2ScSbO_6$ and $k$ = [0, 0.029(1), 0], was suggested in the previous study [39]. However, it is important to note that $Ni_2InSbO_6$ has been reported to exhibit a pronounced ferroelectric polarization below its $T_N$ and multiferroic behavior, which can be attributed to a magnetostriction effect and its inherent helical spin structure [3,39,40]. In contrast, our current study showed that $Ni_2ScSbO_6$ lack any trace of ferroelectric polarization (to be discussed later), thus it is not a multiferroic candidate, unlike $Ni_2InSbO_6$. This fundamental distinction suggests that the spin structure of $Ni_2ScSbO_6$ is essentially different from that of $Ni_2InSbO_6$. Notably, a sinusoidal SDW spin ordering does not allow for ferroelectric polarization, thus ruling out multiferroicity [41]. In this context, it is plausible to attribute the absence of multiferroicity in $Ni_2ScSbO_6$ to the sinusoidal SDW spin ordering, as suggested by our NPD data analysis. However, further investigations using single-crystal neutron diffraction may be beneficial for obtaining a clearer understanding of the spin structure in this system.

During the phase transition from the paramagnetic to an ordered state (viz., AFM) at $T_N$, the intensity of the 003 magnetic Bragg peak can be described by the equation $I(T) = I_{PM} + [I_0(1 - T/T_N)^{2\beta}]$. The first term represents the intensity variation in the paramagnetic region ($T > T_N$), while the second term corresponds to the AFM region ($T < T_N$). Here, $I_0$ is the intensity as $T$ approaches 0 K, and the critical exponent $\beta$ characterizes the nature of the magnetic ordering. Generally, during a phase transition at $T = T_N$, the magnetic order parameter $m \propto \sqrt{I}$. The fitted temperature-dependent integrated intensity of the magnetic reflection (0 0 3) is plotted in Fig. 3 (d). This fit yielded $T_N = 60.6973(4)$ K and a critical exponent $\beta = 0.2198(5)$. The value of $T_N$ is consistent with our bulk magnetization and specific heat measurements. Magnetic systems, depending on their spatial dimensionality and spin characteristics, fall into distinct categories with characteristic $\beta$ values, such as $\beta = 0.125$ for the 2D Ising model, $\beta = 0.33$ for the 3D Ising model, $\beta = 0.35$ for the 3D XY model, and $\beta = 0.36$ for the 3D Heisenberg model. The $\beta$ value of 0.2198(5) for $Ni_2ScSbO_6$ falls within an intermediate regime between the 2D and 3D Ising models. Furthermore, to examine the influence of an external magnetic field on the magnetic ordering, we recorded NPD patterns under magnetic fields up to 4 T at $T = 2$ K. The zero-field NPD data were subtracted from those collected at 4 T, as shown in Fig. S3 (a) of SM. No significant changes indicative of an altered spin structure were observed, demonstrating the robustness of the spin structure under applied magnetic fields, at least up to 4 T.

### E. Dielectric and Magnetodielectric properties

The combination of a polar crystal structure and complex spin ordering in $Ni_2ScSbO_6$ implies potentially interesting dielectric and MD properties. The isostructural compound $Ni_2InSbO_6$ has been reported to exhibit a spin-driven ferroelectric phase transition at its $T_N = 76$ K [26], but these properties are still unexplored in $Ni_2ScSbO_6$. The dielectric constant ($\varepsilon'$) is plotted as a function of temperature under $H = 0$ T for various frequencies ($f$) in Fig. 4(a). The $\varepsilon'(T)$ curves exhibit a λ-shaped peak at $T_E = T_N = 62$ K. The peak position is $f$-independent, whereas the value of $\varepsilon'$ increases over the entire temperature range of interest with increasing $f$; the dielectric loss (tanδ) is negligibly small (~0.0001-0.0003), a characteristic typical of highly insulating materials. These observations imply that extrinsic contributions to $\varepsilon'$ and dielectric relaxation behavior can be ruled out. To investigate whether the peak in $\varepsilon'(T)$ is associated with a ferroelectric phase transition, pyrocurrent *($I_p$)* measurements were conducted both with and without the application of magnetic field. Despite examining multiple samples, no significant $I_p$ signal was detected. Our SXRD study (see below) reveals a significant magnetostriction effect that is likely associated with the $\varepsilon'(T)$ peak. Although magnetostriction can induce ferroelectric polarization *(P)* in various systems, leading to multiferroicity [27,42], the absence of an $I_p$ signal in $Ni_2ScSbO_6$ apparently suggests a distinctly different nature to the isostructural compound $Ni_2InSbO_6$.

To investigate whether $Ni_2ScSbO_6$ exhibits MD coupling, $\varepsilon'(T)$ curves were measured under the application of various magnetic fields, as shown in Fig. 4(b). The $\varepsilon'(T)$ peak is suppressed significantly with increasing *H*. Moreover, the peak shifts towards lower *T*, which can be ascribed to the typical suppression of $T_N$ with field for AFM systems. According to the Lyddane-Sachs-Teller (LST) relation, the dielectric permittivity can be inherently coupled to the optical phonons, which can be expressed as $\frac{\varepsilon_0}{\varepsilon_\infty} = \prod_j \frac{\omega_{LOj}^2}{\omega_{TOj}^2}$ ; here $\varepsilon_0$ refers to the static dielectric constant, $\varepsilon_\infty$ represents the optical dielectric constant, $\omega_{LO}$ and $\omega_{TO}$ denote the long-wavelength longitudinal (LO) and transverse (TO) optical-phonon mode frequencies, respectively [34,43]. Generally, $\omega_{LO}$ is a *T*-independent parameter, whereas $\omega_{TO}$ is closely linked to the $\varepsilon'(T)$ curves. According to the modified Barrett theory, this relation is expressed as $\varepsilon'(T) = \varepsilon'(0) + A/[\exp\left(\frac{\hbar\omega_0}{k_B T}\right) - 1]$, where $A$ denotes a coupling constant and $\omega_0$ represents the average frequency of the low-lying phonon modes [43]. The $\varepsilon'(T)$ curves of $Ni_2ScSbO_6$ closely follow the Barrett theory above $T_N$, as shown in the inset of Fig. 4(b). However, $\varepsilon'(T)$ deviates from the Barrett curve near $T_N$, thereby indicating MD coupling. The fit parameters obtained are $\varepsilon'(0) = 7.283, A = 0.531$ and $\omega_0$ =291.38 cm$^{-1}$. The obtained value of $\omega_0$ is consistent with the observed low-lying Raman mode frequencies (see below). A similar scenario has been observed in other systems exhibiting the MD effect, which has been attributed to the magnetostriction effect combined with spin-phonon coupling (SPC) [14,42,44]. $Ni_2ScSbO_6$ exhibits a pronounced SPC effect (to be discussed later) below $T_N$, which may be a plausible origin for the MD coupling in this system.

The isothermal dependence of $\varepsilon'$ at $T$ = 2 K is shown in Fig. 4 (c). With increasing *H*, a distinct and systematic decrease in $\varepsilon'$ is observed. The associated MD coupling parameter (*MD*%) $=[\varepsilon'(0)- \varepsilon'(H)/ \varepsilon'(0)]\times 100\%$ was calculated, and is shown in the inset of Fig. 4 (c). This further confirms the occurrence of a pronounced MD coupling in this system.

## F. Temperature-dependent Raman spectroscopy: Spin-phonon coupling

Raman spectroscopy stands out for its unique ability to pinpoint structural alterations or distortions on a local scale, alongside gauging electronic excitations, the degree of cationic ordering, and SPC in various systems. In the case of $Ni_2ScSbO_6$, the potential existence of SPC was hinted at by the deviation of $\varepsilon'(T)$ from the LST relation near $T_N$, as described above. Therefore, to seek direct evidence of SPC and explore the phonon behavior in $Ni_2ScSbO_6$, especially around $T_N$ = 62 K, a temperature-dependent Raman spectroscopy study was conducted, as illustrated in Figs. 5 (a)-(c). The spectra show no signals of structural phase transitions within the studied temperature range, as the mode distribution remains consistent down to 10 K, aligning with SXRD results discussed below. This aligns with the results obtained from SXRD (discussed below). However, a relatively broad peak centered at $\omega$ = 159 cm$^{-1}$ emerges below 150 K. In the absence of any structural phase transitions within this temperature range, the peak may be associated with a magnon (M) mode, which is often observed at temperatures well above the long-range magnetic ordering temperatures [45–48]. Moreover, a relatively large blueshift ($\Delta\omega$ ~ 6 cm$^{-1}$) of the suspected magnon peak in $Ni_2ScSbO_6$ is observed as $T$ is decreased from 150 to 10 K, as shown in the SM, Fig. S6. A similar large blueshift was reported for a magnon peak pertaining to the AFM system $FePS_3$ [49]. The temperature variation ($T$-variation) of the linewidth $(\Gamma)$ of this possible magnon mode (M) closely follows $T^{\alpha}$ behavior below $T_N$, where $\alpha \approx 4.06$ (inset of SM Fig S6) [31]. In fact, the linewidth of a magnon mode within a small $T$-regime is predicted to follow $T^{4}$ behavior according to the Hartree-Fock approximation [50]. To confirm the magnon mode unambiguously, further investigation of this peak under a magnetic field may be beneficial.

Utilizing a group theory-based symmetry analysis on the $Ni_2ScSbO_6$ structure, characterized by trigonal symmetry $R3$ (no. 146) with a point group $C_3$ (3), predicts a total of 30 possible modes at the Γ-point of the Brillouin zone. The phonons exhibiting infrared (IR) and Raman responses are predicted to consist of 27 simultaneously active Raman ($9\,A + 9\,{}^1E + 9\,{}^2E$) and IR ($9\,A + 9\,{}^1E + 9\,{}^2E$) modes, along with three acoustic modes ($A + {}^1E + {}^2E$). Here, the symmetry $E$ corresponds to a doubly degenerate mode. The Raman peaks observed are labeled as P1-P23 in Fig. 5, and detailed information regarding the tentative symmetry assignments is summarized in Table SII of the SM [31].

In general, for a magnetic system, the $T$-variation of a phonon mode frequency $\omega$ is described as: $\omega(T) = \omega_0 + \Delta\omega_{latt}(T) + \Delta\omega_{anh}(T) + \Delta\omega_{e\text{-}ph}(T) + \Delta\omega_{s\text{-}ph}(T)$, where $\omega_0$ represents the frequency at an initial temperature $T_0$, $\Delta\omega_{latt}$ arises from the contraction or expansion of the crystal lattice, $\Delta\omega_{anh}$ stands for the anharmonic contribution, and $\Delta\omega_{e\text{-}ph}$ and $\Delta\omega_{s\text{-}ph}$ represent the changes originating from electron-phonon and spin-phonon couplings, respectively. $Ni_2ScSbO_6$ is highly insulating in nature, thus any contribution from $\Delta\omega_{e\text{-}ph}$ can be neglected. Our analysis shows that the integrated peak intensities *($I_R$)* of various Raman peaks exhibit anomalies at $T_N$, thus signaling a possible correlation between spins and phonons in this system as shown in SM. In the absence of spin–phonon coupling and any structural changes or lattice distortions, the temperature dependence of $\omega$ can be described solely by the anharmonic behavior outlined in Balkanski's model: $\omega_{anh}(T) = \omega_0 - C\left(1 + \frac{2}{e^{\frac{\hbar\omega_0}{2k_BT}}-1}\right)$; here, $C$ is a variable parameter, $\hbar$ denotes the reduced Planck's constant, and $k_B$ signifies the Boltzmann constant [51]. Thus, a phonon mode exhibiting anharmonic behavior is anticipated to exhibit a low-$T$ plateau in frequency, succeeded by a linear decrease at elevated temperatures. The Raman modes of $Ni_2ScSbO_6$ with high or moderate intensities were analyzed using Lorentzian spectral decomposition. Anharmonic fitting was performed on these curves (shown in red) in the elevated $T$ region ($T > 60$ K). Notably, its several modes deviate from the expected anharmonic behavior, as shown in SM, Fig. S8. Figure 5(c) depicts the $\omega(T)$ curve for the P10 mode with the maximum deviation. The pronounced degree of anomalous softening observed below $T_N$ is a typical signature of SPC [1,14,42,52–54]. For some modes, the deviation of $\omega(T)$ from the anharmonic model begins above $T_N$, which is attributable to significant short-range magnetic ordering, as evidenced by our magnetization and specific heat studies. In the magnetically ordered state of a system, the development of strong spin correlations can cause phonon renormalization, which can be described by a formalism proposed by Granado *et. al*. According to this formalism, [55] the relation between the nearest-neighbor spin-spin correlation function $\langle \boldsymbol{S_i}.\boldsymbol{S_j}\rangle$ and the change in the phonon frequency ($\Delta\omega_{s-ph}$) can be expressed as: $\Delta\omega_{s-ph}(T) \approx \lambda\langle \boldsymbol{S_i}.\boldsymbol{S_j}\rangle \approx \lambda\left\{\frac{M(T)}{M_{Max}}\right\}^2$; wherein $\lambda$ is a constant measuring the strength of the SPC and $M(T)/M_{Max}$ refers to the $T$ variation of the normalized $M$ curve with respect to maximum value of $M$, i.e. $M_{Max}$. Plots of $\Delta\omega$ and $\left\{\frac{M}{M_{Max}}\right\}^2$ versus $T$ are shown for modes $P10$ and $P22$ in Figs. 7 (a), (b), which display matching trends from $T_N$ down to $T = 30$ K. A similar result is found for the other modes with anomalous phonon softening, as shown in Fig. S9 of the *SM* [31]. This clearly confirms the presence of SPC. It is relevant here to mention that a frequency-dependent peak in the *ac* susceptibility [$\chi''(T)$] data was observed near $T = 30$ K, which is likely attributed to a reentrant glassy state associated with the partial freezing of spins in this temperature range (see Fig. S1 in the *SM*) [34]. This is possibly the reason why the phonon renormalization does not follow $\left\{\frac{M}{M_{Max}}\right\}^2$ behavior at lower temperatures. Similar to $\omega(T)$, the corresponding $\Gamma(T)$ curves can be described by an anharmonic relation expressed as follows: $\Gamma_{anh}(T) = \Gamma_0 + A\left(1 + \frac{2}{e^{\frac{\hbar\omega_0}{2k_BT}}-1}\right)$; here $A$ is an adjustable parameter, and $\Gamma_0$ represents $\Gamma$ as $T \rightarrow 0$ K [42,51]. The $\Gamma(T)$ curves of $P10$ and $P22$ Raman modes, along with their anharmonic fits (shown in red), are presented in the insets of Figs. 6

(a) and (b), respectively. The *Γ(T)* curves for the other analyzed modes are shown in Fig. S10 of the SM [31] The phonon lifetime ($\tau$) is directly related to $\Gamma$ as $\tau = 1/(2\pi c \times \Gamma)$. An anomaly in $\omega(T)$ may also be driven by magnetostriction effect, but this typically does not affect $\tau$ and thus should not influence the *Γ(T)* curves. The majority of the modes show a clear deviation in *Γ(T)* from the anharmonic model below $T_N$. This, provides unambiguous confirmation of the occurrence of SPC in this system. Furthermore, the $\lambda$ values for various modes were estimated from linear fits to the $\Delta\omega\ vs. \left\{\frac{M(T)}{M_{Max}}\right\}^2$ plots shown in Figs. 7 (c), (d) for the P10 and P22 modes; fits for the other analyzed modes are shown in Fig. S11 in the SM [31]. The obtained λ values range from 0.275 to 1.03 $cm^{-1}$, which can be compared with values reported for previously studied AFM compounds exhibiting SPC effects, such as $MnF_2$ ($\lambda$ = 0.4 $cm^{-1}$), $CuB_2O_4$ ($\lambda$ = 0.02−0.03 $cm^{-1}$), $Ni_2InSbO_6$ ($\lambda$ = 0.55−2.14 $cm^{-1}$), $Ni_2NbBO_6$ ($\lambda$ = − 0.3 to + 0.67 $cm^{-1}$), $ZnCr_2O_4$ ($\lambda$ = 3.2−10.0 $cm^{-1}$), $CrSiTe_3$ ($\lambda$ = 0.1−0.2 $cm^{-1}$), and 15*R*- $BaMnO_3$ ($\lambda$ =1.2−3.8 $cm^{-1}$) [45,56–61]. This suggests that the SPC effect occurring in $Ni_2ScSbO_6$ is relatively strong. A theoretical estimation of $\lambda$ for the P10 mode (showing the most prominent deviation) was performed following the simplified lattice model of Sushkov *et al.* [61]: $\lambda \approx \frac{2\alpha^2 J}{m\omega_0}$, where $\alpha = 2z/3a_B = 6.301$ Å$^{-1}$ (here, $z$ refers to the nearest-neighbor co-ordination number of Ni, $a_B$ denotes the Bohr radius, $J$ represents the exchange coupling constant, $\omega_0 = 2\left(\frac{K0}{m}\right)^{1/2}$ denotes the frequency of the phonon mode as calculated from the Raman shift, and $m$ is the mass of a Ni atom. Here, $J = 3k_B\theta_{CW}/zS(S+1)$ = 1.87 meV, where $\theta_{CW}$ = – 87 K, $k_B$ denotes the Boltzmann constant, and $S$ = 1. The calculation results in $\lambda \sim 1.9$ $cm^{-1}$, which is consistent with the experimentally obtained value of $\lambda \sim 1.03$ $cm^{-1}$.

### G. Synchrotron x-ray diffraction: magnetostriction effect

$\Delta\omega_{latt}$, arising from changes in ionic binding energy induced by the lattice contraction or expansion, may contribute significantly to the changes in the phonon frequencies [55]. Hence, a comprehensive study of crystal structure evolution across magnetic transition temperatures in SPC systems is essential. We performed high-resolution SXRD measurements from 300 K to 20 K, as shown in Fig. S12(a) and (b) of the SM. No peak splitting and/or drastic changes in peak shapes were observed throughout the studied temperature range, ruling out a structural phase transition down to 20 K. All data were satisfactorily fitted using the same trigonal symmetry structural model (*R*3). The SXRD data collected at $T$ = 100 K and 20 K, along with their Rietveld refined profiles, are shown in Fig. S13 of the *SM* [31]. However, integrated peak intensity analysis revealed an anomalous change at $T_N$ for various Bragg peaks, as demonstrated in Fig. S14 of the *SM* [31]. Further investigations into the thermal variation of the lattice parameters ($a$ = $b$, $c$), and the unit cell volume ($V$) reveal pronounced anomalies at $T_N$, as evident in Figs. 8 (a), (b), and (c). A dramatic down-turn is observed in *a(T)* below $T_N$, indicating an anomalous lattice contraction along the crystallographic *a*-axis. In contrast, the *c(T)* curve exhibits a pronounced up-turn below $T_N$, indicating anomalous thermal expansion along the *c*-axis. A distinct down-turn in the *V(T)* curve is also observed below $T_N$, indicating a sizeable lattice distortion. If the lattice distortion is sufficiently large, quantifying the spontaneous volume magnetostriction effect using this method is advantageous compared to strain gauge and capacitive techniques [62,63]. The thermal expansion law can be applied to the lattice parameters using the following expressions: $a(T) = a_0\left[1 + \frac{Be^{\frac{d}{T}}}{T\left(e^{\frac{d}{T}}-1\right)^2}\right]$; $c(T) = c_0\left[1 + \frac{fe^{\frac{g}{T}}}{T\left(e^{\frac{g}{T}}-1\right)^2}\right]$; and $V(T) = V_0\left[1 + \frac{A}{\left(e^{\frac{\theta_D}{T}}-1\right)}\right]$; here $a_0$, $c_0$, and $V_0$ denote the lattice constants at $T$ = 0 K, whereas *B*, *d*, *f*, *g*, and *A* denote adjustable parameters. $\theta_D$ refers to the Debye temperature [14]. While the *T*-dependence of the lattice parameters closely follows the expected thermal expansion behavior above $T_N$, a pronounced deviation is observed below $T_N$, as illustrated in Figs. 8 (a)-(c). This deviation suggests an isostructural distortion resulting from a magnetostriction effect, which is induced by the onset of complex AFM spin ordering below $T_N$. The various fitting parameters thus obtained are summarized in Table SIII of the *SM* [31]. The spontaneous volume magnetostriction ($\lambda_{ms}^V$) at a specific temperature can be estimated by examining the difference in volume between a system in an ordered magnetic state and in the paramagnetic (PM) state [64] as expressed by the formula: $\lambda_{ms}^V(T) = \frac{V_{AFM}(T)-V_{PM}(T)}{V_{PM}(T)}$. Here, $V_{AFM}(T)$ denotes the volume of the system in the AFM state at a temperature *T*, while $V_{PM}(T)$ refers to the volume the system would have if it were in the PM state at the same temperature. For $Ni_2ScSbO_6$, the inset of Fig. 8 (c) displays $\lambda_{ms}^V$ as a function of *T*, highlighting the significant emergence of the magnetostriction effect below $T_N$. At $T$ = 20 K, $\lambda_{ms}^V$ is ~3 × $10^{-4}$, which is comparable to the spontaneous volume magnetostriction values previously reported for 15*R*- $BaMnO_3$ ($\lambda_{ms}^V$ ~ 32×$10^{-4}$ at 90 K), $Fe_3(PO_4)O_3$ ($\lambda_{ms}^V$ ~ 7.8×$10^{-4}$ at 100 K), and $Zn_{1-x}Cu_xCr_2Se_4$ ($\lambda_{ms}^V$~ 4.6–24.9×$10^{-4}$ at 100 K for varying x values) [14,42,64].

In addition to the observed changes in the lattice constants *a*, *c*, and *V*, our analysis suggests that the $Ni(1)O_6$ and $Ni(3)O_6$ octahedra showed concomitant changes in their associated bond-lengths and bond-angles below $T_N$., as demonstrated in Fig. 9 (a) -(f). Additional parameters are shown in Fig. S15 of the SM [31]. The corresponding structural distortion below $T_N$ is schematically illustrated in Fig. 9 (g). Nearest neighbor Ni spins (with a distance ~ 3.78 Å) are coupled via the superexchange pathway Ni(1)-O(1)-Ni(3). Next-nearest neighbor Ni spins (Ni(1)-Ni(1) or Ni(3)-Ni(3) pairs lying in the *ab* plane at a distance ~ 5.16 Å) are coupled through weaker exchange pathways via $ScO_6$ and $SbO_6$ involving both O(1) and O(2). Our SXRD analysis indicates a much larger displacement for O(1) than for O(2) [as evident from Fig. 9 (g)]. This can presumably be attributed to the predominant exchange interactions via the Ni(1)-O(1)-Ni(3) exchange pathway.

The Grüneisen model connects changes in phonon mode frequency $(\Delta\omega/\omega)$ to changes in unit-cell volume $(\Delta V/V)$ [14]. This relationship is expressed as $(\Delta\omega/\omega) = -\gamma(\Delta V/V)$, where $\gamma$ is the Grüneisen parameter for a given phonon mode. Plots of $(\Delta\omega/\omega)$ *vs.* $(\Delta V/V)$ exhibit a linear nature, as illustrated in Figs. 10 (a) and (b) for modes P10 and P22 and Fig. S16 of the SM for other relevant modes. This further emphasizes the crucial role of lattice distortion in inducing anomalous phonon behavior in $Ni_2ScSbO_6$. Consequently, the present study confirms strong direct spin-phonon coupling, along with lattice distortion driven by the magnetostriction effect that facilitates indirect coupling of spins and phonons. Typically, the contribution of magnetostriction to the SPC effect is considered negligible when reporting SPC effects in various systems [65,66]. Thus, the present study may provide new insights into the SPC effect, particularly in systems with complex magnetic ordering.

## IV. CONCLUSION

In conclusion, our comprehensive study of the intriguing material $Ni_2ScSbO_6$ reveals a range of exotic phenomena that are intricately interconnected due to cross-coupling among spin, dipole, lattice, and phonon degrees of freedom. We observe incommensurate spin-density wave antiferromagnetic ordering at $T_N = 62$ K, which remains robust under magnetic fields of up to 4 T, as demonstrated by magnetization, specific heat, and neutron diffraction studies. Dielectric measurements exhibit a λ-shaped peak in the $\varepsilon'(T)$ curves, with its position remaining unchanged across varying frequencies, while its magnitude is significantly suppressed under an applied magnetic field, confirming pronounced magnetodielectric coupling at $T_N$. Unlike isostructural ferroelectric materials $Ni_2InSbO_6$ and $Ni_3TeO_6$, $Ni_2ScSbO_6$ lacks ferroelectric ordering, attributable to its distinctive sinusoidal spin-density wave magnetic ordering that preserves inversion symmetry. The correspondence between spin and phonon behavior is illustrated by the anomalous softening of various Raman modes. Furthermore, high-resolution synchrotron X-ray diffraction studies quantitatively reveal a substantial magnetostriction effect, characterized by an unusual thermal variation in the unit cell volume $(V)$ below $T_N$. Unlike conventional spin-phonon coupling phenomena, where the magnetostriction effect remains minimal or absent, in this case, the observed phonon modulation is significantly influenced by both direct spin-phonon coupling and the magnetostriction effect, as validated by Grüneisen's law. The strong correlations among these phenomena provide valuable insights into the coupling mechanisms of various microscopic degrees of freedom in other complex magnetic systems.

**ACKNOWLEDGMENTS**

SC is thankful to CIFC, IIT (BHU) Varanasi, for providing access to the magnetometer (MPMS) facility. HDY acknowledges the generous support from the National Science and Technology Council of Taiwan through the grants: NSTC-113-2112-M110-006 and NSTC-113-2112-M110-022. A.P. gratefully acknowledges the support received from the European Commission through the Horizon Europe MSCA Postdoctoral Fellowship, project number 101110742. We also express our sincere gratitude to the Australian Nuclear Science and Technology Organization (ANSTO) for granting essential access to their experimental facilities.

**Table I**: Neutron powder diffraction (NPD) fit parameters for different models of magnetic ordering.

| **Model** | $\chi^2$ | **Magnetic R-factor** | ***M*** **($\mu_B$)** |
|---|---|---|---|
| Model_1: SDW | 13.1 | 5.870 | 2.7261 |
| Model_2: Spins in *ac* plane | 13.5 | 7.550 | 2.6488 |
| Model_3: Spins in *bc* plane | 13.8 | 7.527 | 2.6672 |
| Model_4: Spins in *ab* plane | 20.2 | 22.49 | 1.6108 |

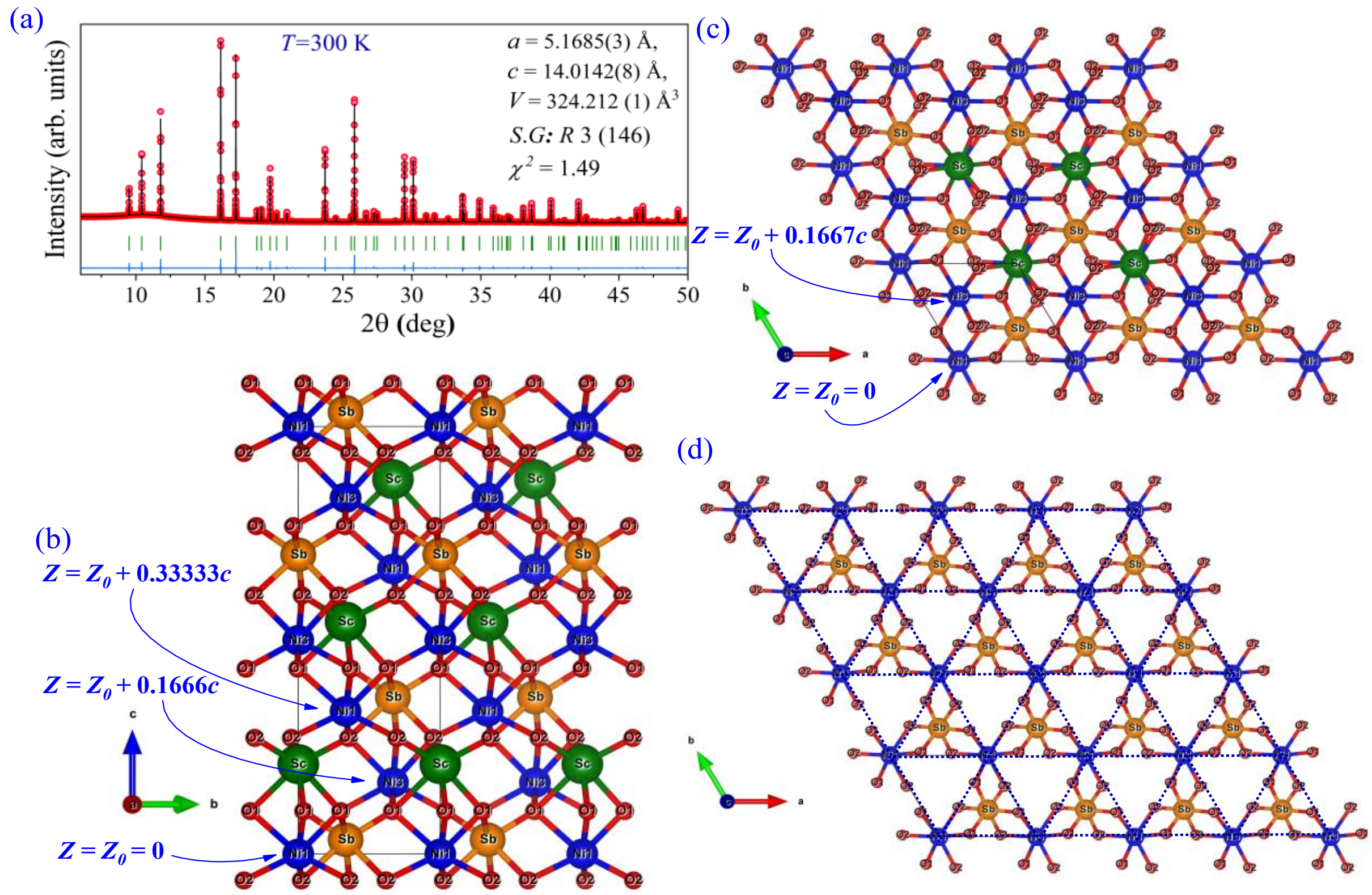

Fig. 1. (a) XRD pattern of $Ni_2ScSbO_6$ at 300 K (red data points) fitted by Rietveld refinement (black line). The difference profile is shown in blue. (b) Representation of the three-dimensional crystal structure. (c) and (d) Views of the crystal structure in two different *ab* planes, on which the Ni atoms form triangular networks. The blue dashed lines serve as a guide to the eye.

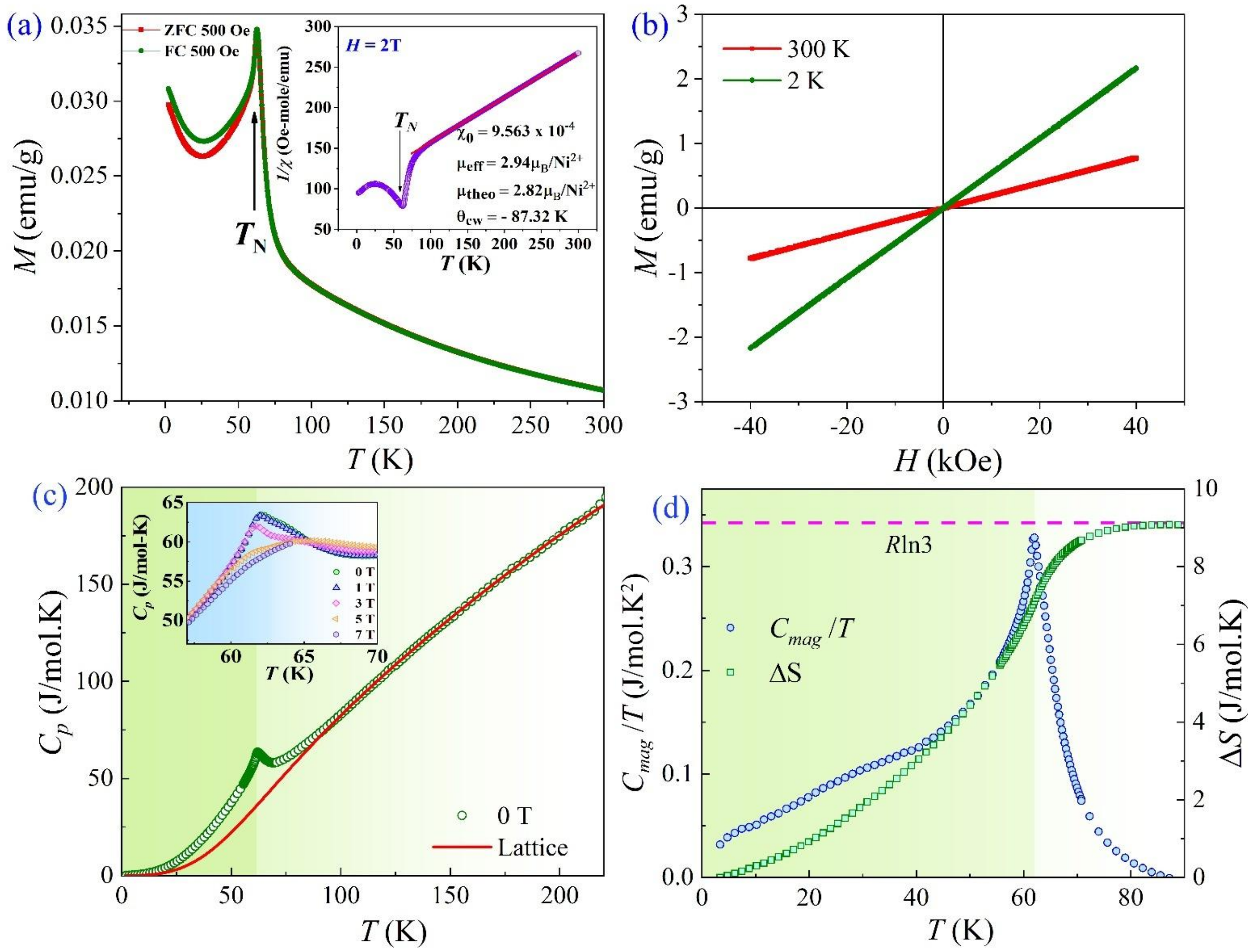

Fig. 2. (a) *T*-dependence of ZFC and FC magnetic susceptibility recorded under $H$ = 500 Oe. Inset: Curie-Weiss fit of $1/\chi$ *vs.* $T$ curve (recorded under $H$ = 20 kOe) in the paramagnetic regime. (b) Isothermal magnetization versus applied field curves measured at $T$ = 2 and 300 K. (c) Temperature-dependent zero-field specific heat ($C_p$) curve along with its Debye function fit (shown in red). Inset: $C_p(T)$ curves recorded under various $H$. (d) $C_{mag}/T$ *vs.* $T$ curve at $H$ = 0 T and the corresponding magnetic entropy $(\Delta S)$ variation as a function of temperature. Here, $C_{mag}$ denotes the magnetic specific heat.

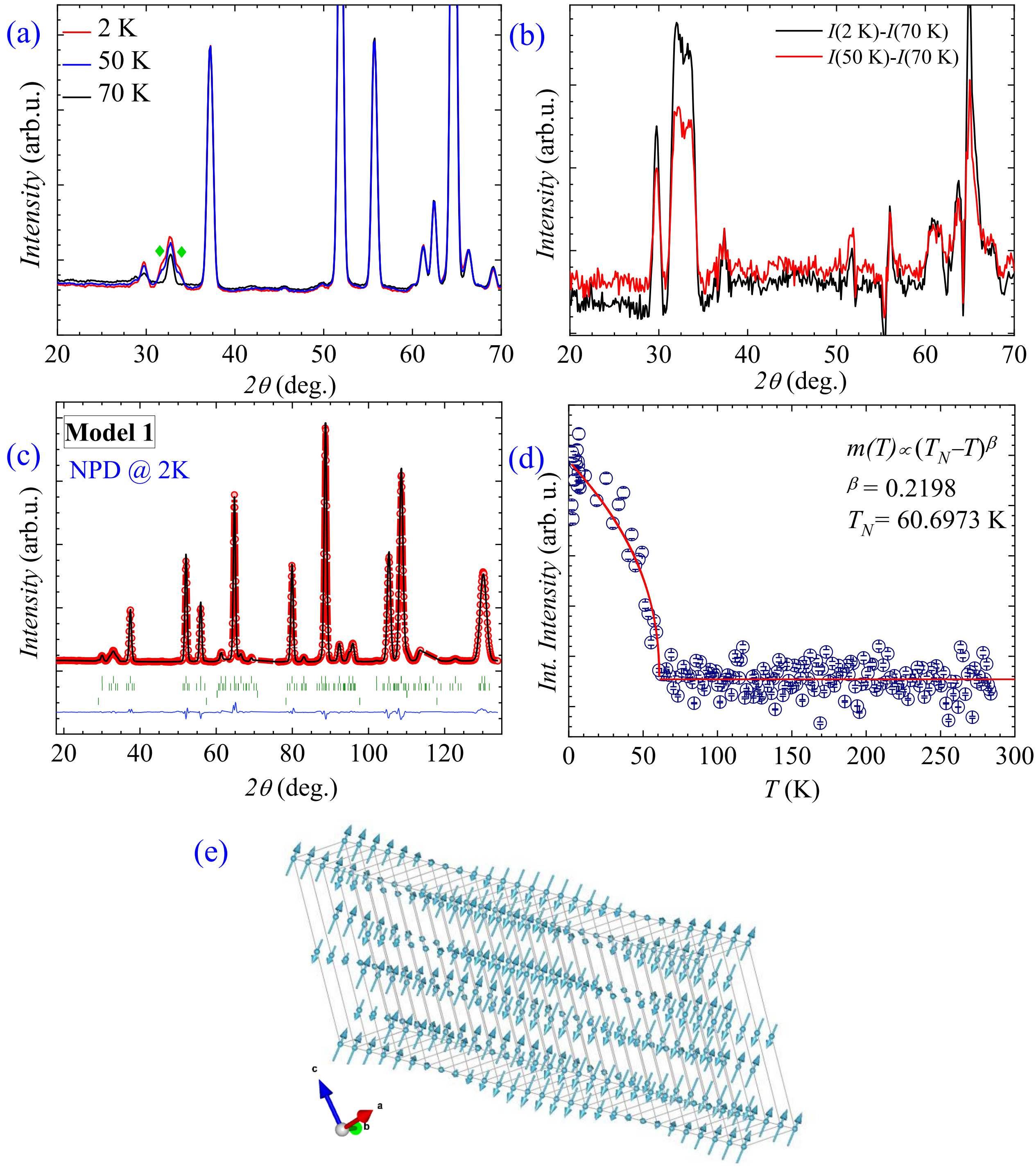

Fig. 3. (a) Neutron diffraction patterns at 2 K, 50 K, and 70 K. The peaks marked by green diamond symbols indicate purely magnetic peaks that appear only below $T_N$. (b) Neutron diffraction patterns at 2 K and 50 K with the 70 K data subtracted to show the magnetic Bragg peaks and (c) Observed (red data points), fitted (black line) and difference (blue line) neutron diffraction profiles at 2 K. The green vertical bars represent (top to bottom) Bragg reflection positions for $Ni_2ScSbO_6$ (nuclear and magnetic) and NiO (nuclear and magnetic) impurity, respectively. (d) Power law fit to the temperature dependence of the integrated intensity of the (003) Bragg reflection. (e) A pictorial representation of the SDW spin ordering of $Ni_2ScSbO_6$ as obtained from the analysis of the NPD data at 2 K.

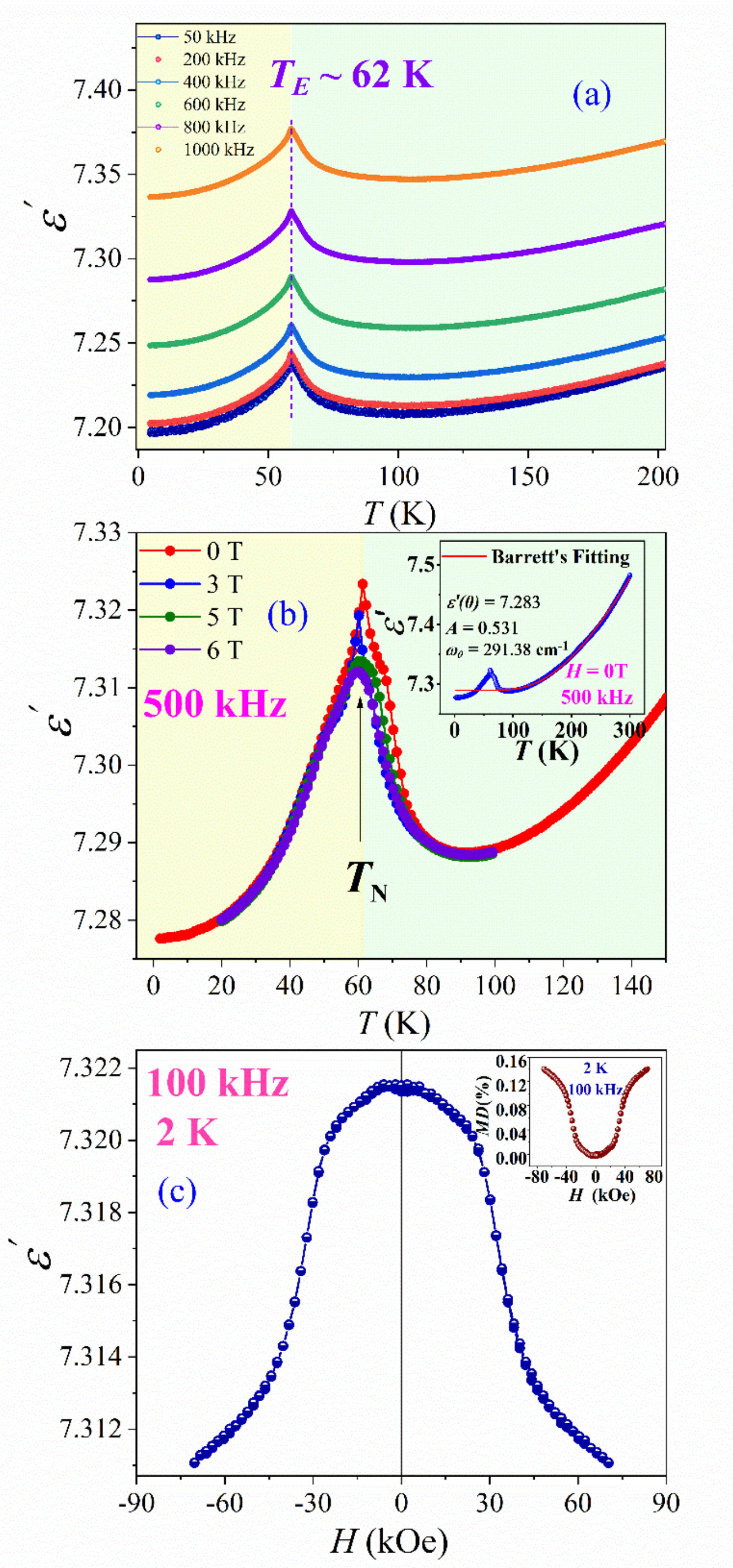

Fig. 4. (a) Temperature-dependence of dielectric constant ($\varepsilon'$) at various frequencies. (b) $\varepsilon'(T)$ curves measured under various magnetic fields. Inset: Barrett's fit to $\varepsilon'(T)$ curve measured with a frequency of 500 kHz and under $H$ = 0 T. (c) Isothermal magnetic field variation of $\varepsilon'$ at 2 K and $f$ = 100 kHz. Inset: Field variation of magnetodielectric coupling parameter (MD%).

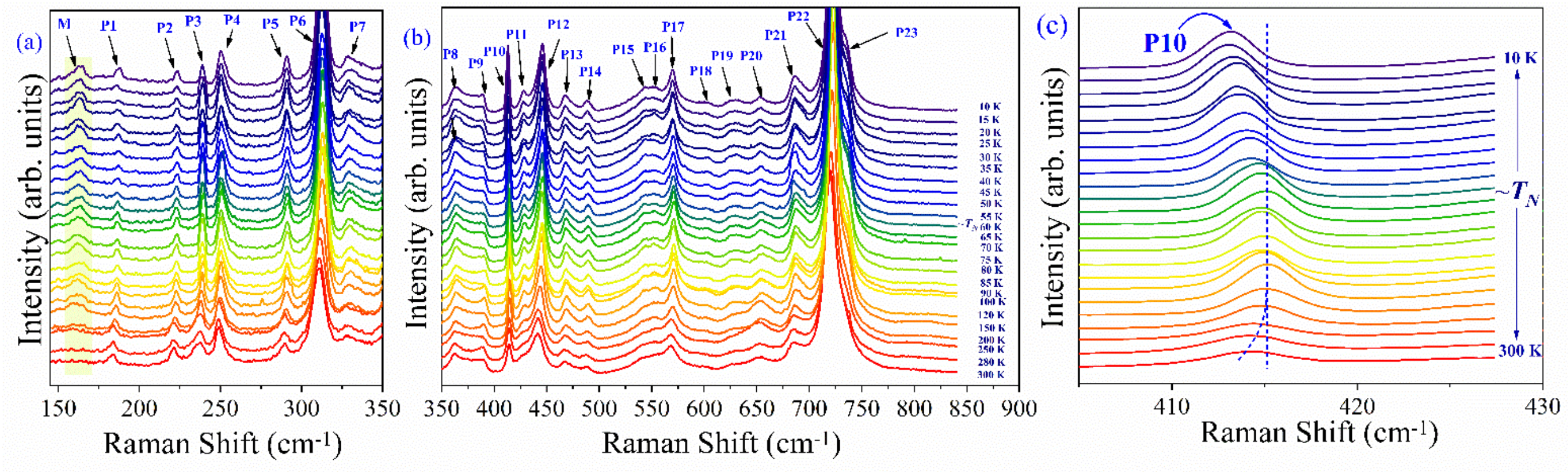

Fig. 5. (a) and (b) Raman spectra in two different frequency regions, collected at various temperatures. Corresponding phonon modes are designated as $P1$ to $P23$. The mode M appears only at lower temperatures, and is denoted as a possible magnon mode. (c) Demonstrates the anomalous softening of the mode P10. The blue dashed lines are the guides to the eye.

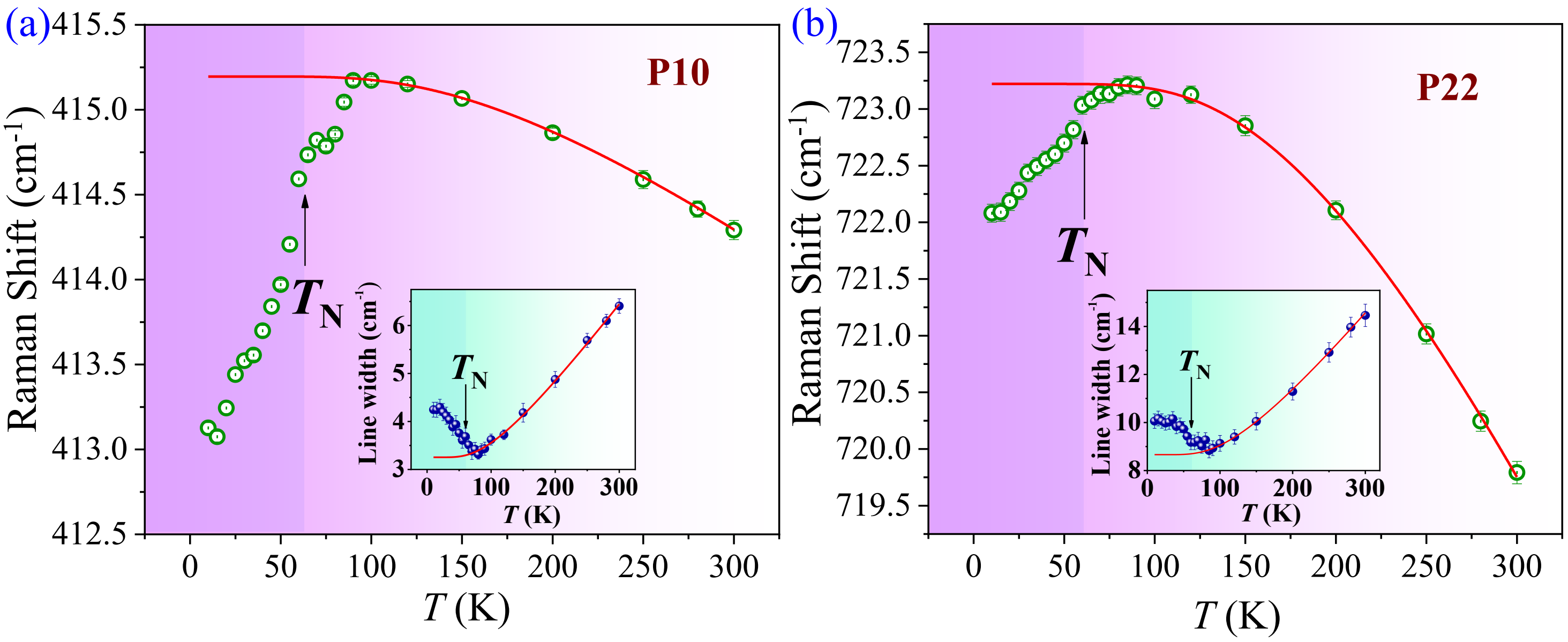

Fig. 6 (a) and (b) Temperature dependence of P10 and P22 phonon mode frequencies, with experimental data (green circles) fitted using an anharmonic law (red solid curve). Insets display the corresponding FWHM dependence on temperature.

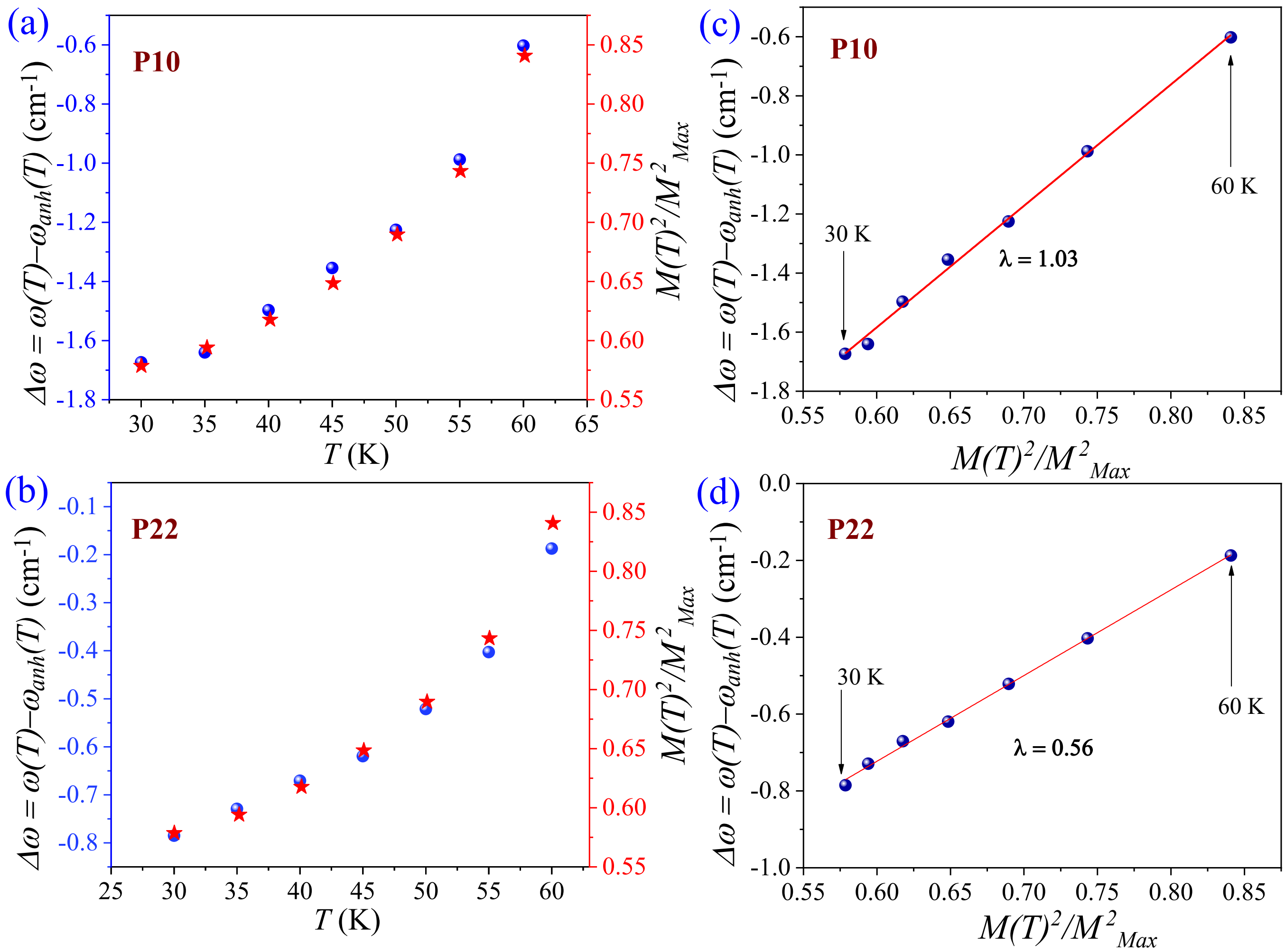

Fig. 7. (a) and (b) Temperature dependence of $\Delta\omega$ and $\{M(T)/M_{Max}\}^2$ for the P10 and P22 modes, respectively. (c) and (d) $\Delta\omega$ *vs.* $\{M(T)/M_{Max}\}^2$ plots for P10 and P22, with linear fits shown by the red lines.

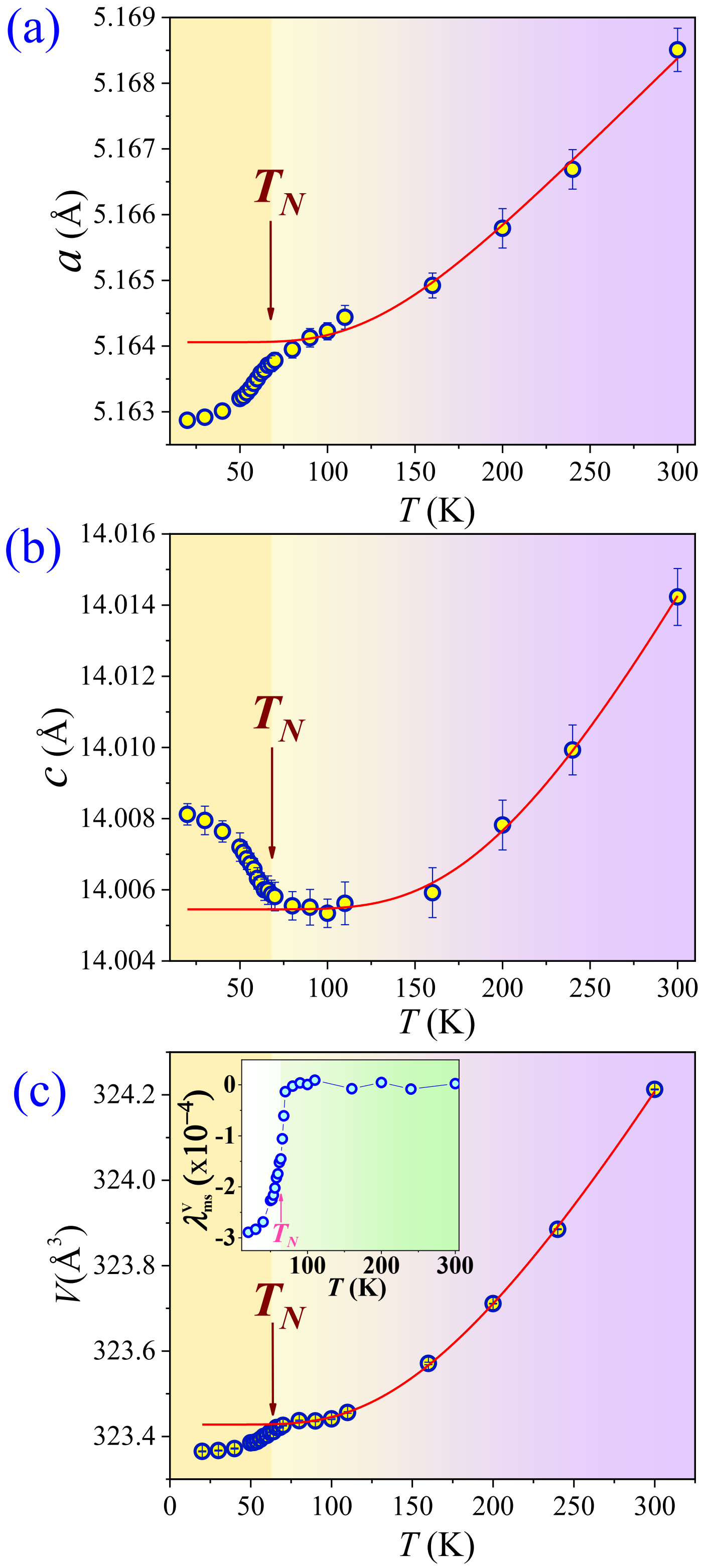

Fig. 8. (a), (b) and (c) Temperature dependence of lattice parameters *a*, *c*, and unit-cell volume *(V)*, along with fits to thermal expansion model (red line) in the paramagnetic region. The inset of (c) shows the volume magnetostriction coefficient as a function of temperature.

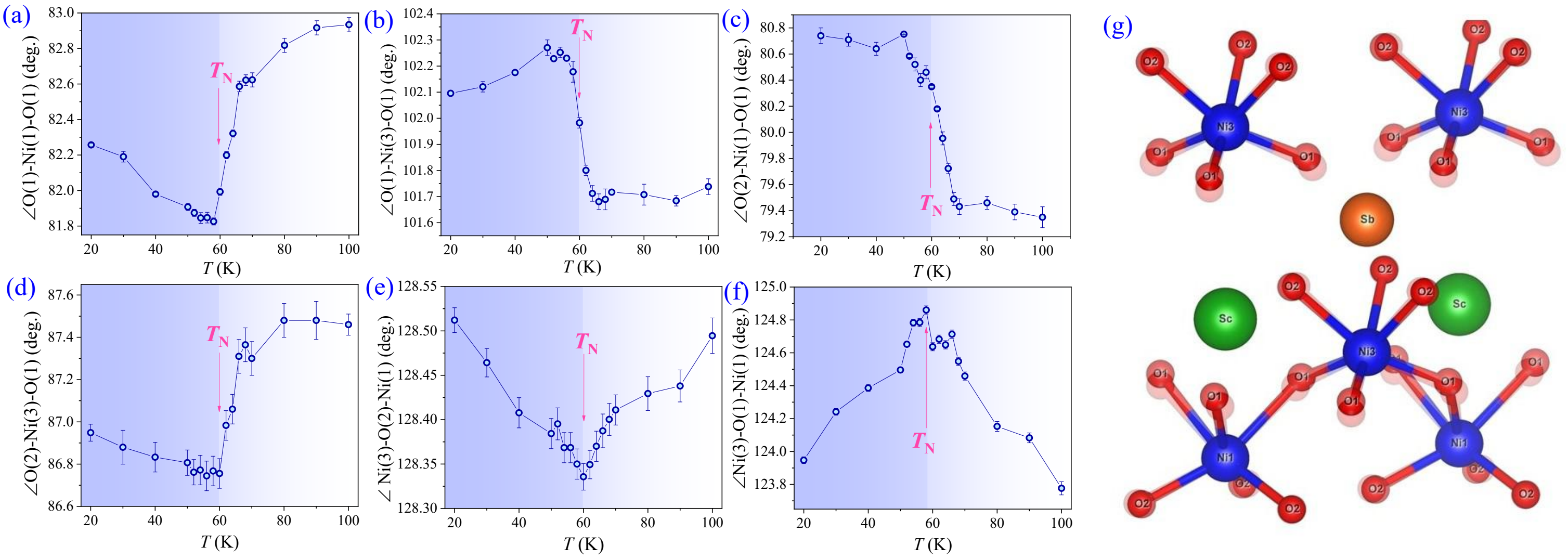

Fig. 9 (a)-(f) Temperature dependence of various bond-angles related to $Ni(1)O_6$ and $Ni(3)O_6$ octahedra. $T_N$, (g) Depiction of change in octahedral structure due to magnetostriction, where the shadowed and solid red spheres represent the positions of oxygen atoms above and below $T_N$.

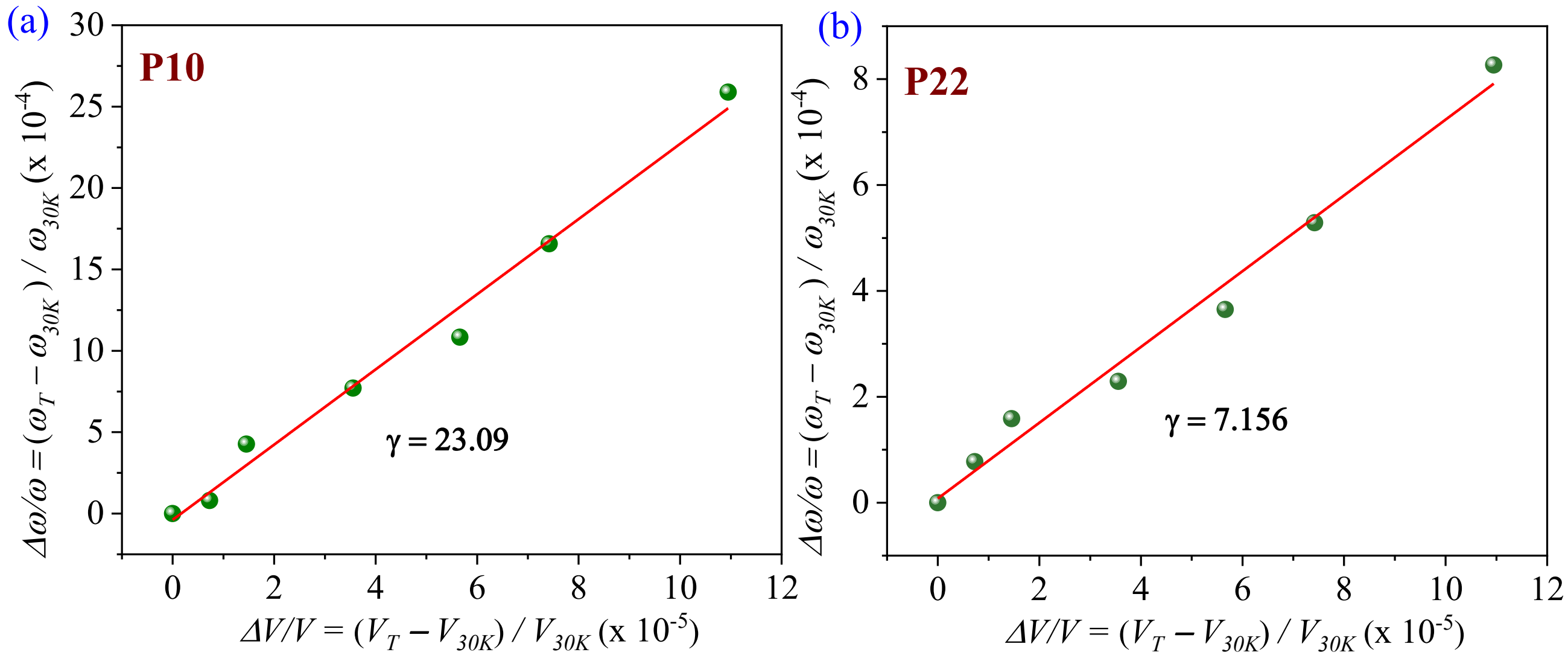

Fig. 10. (a) and (b) *(Δω/ω) vs. (ΔV/V)* curves for P10 and P22 modes, respectively. The Grüneisen parameter, $\gamma$, derived from linear fits (shown in red), indicates a strong correlation between the lattice distortion and phonons.

**Supplementary**

**Table SI**: Refined structural parameters obtained from fitting the SXRD data collected at 300 K, 100 K and 20 K.

Space group *R*3(146)

| Parameters | *T* = 300 K | *T* = 100 K | *T* = 20 K |
|---|---|---|---|
| **$\lambda$ (Å)** | 0.774890 | 0.619920 | 0.619920 |
| ***a* (Å)** | 5.168508(8) | 5.164224(13) | 5.162872(10) |
| ***b* (Å)** | 5.168508(8) | 5.164224(13) | 5.162872(10) |
| ***c* (Å)** | 14.01423(3) | 14.00535(4) | 14.00812(3) |
| ***V* ($Å^3$)** | 324.2129(9) | 323.4705(15) | 323.3651(12) |
| **α** | 90.00 | 90.00 | 90.00 |
| **β** | 90.00 | 90.00 | 90.00 |
| **γ** | 120.00 | 120.00 | 120.00 |
| **Ni1** | | | |
| **x** | 0.00000 | 0.00000 | 0.00000 |
| **y** | 0.00000 | 0.00000 | 0.00000 |
| **z** | 0.00000 | 0.00(8) | 0.00(8) |
| **Sc** | | | |
| **x** | 0.00000 | 0.00000 | 0.00000 |
| **y** | 0.00000 | 0.00000 | 0.00000 |
| **z** | 0.21000(8) | 0.21000(8) | 0.21000(8) |
| **Ni3** | | | |
| **x** | 0.00000 | 0.00000 | 0.00000 |
| **y** | 0.00000 | 0.00000 | 0.00000 |
| **z** | 0.50000 | 0.50000 | 0.50000 |
| **Sb** | | | |
| **x** | 0.00000 | 0.00000 | 0.00000 |
| **y** | 0.00000 | 0.00000 | 0.00000 |
| **z** | 0.70000(8) | 0.70000(8) | 0.70000(8) |
| **O1** | | | |
| **x** | 0.33670(8) | 0.33484(6) | 0.32725(6) |
| **y** | 0.03800(10) | 0.04184(8) | 0.03470(10) |
| **z** | 0.10010(10) | 0.09840(10) | 0.09964(8) |
| **O2** | | | |
| **x** | 0.30600(6) | 0.31275(6) | 0.31506(6) |
| **y** | 0.35480(8) | 0.36200(8) | 0.3612(6) |
| **z** | 0.2713(4) | 0.27151(2) | 0.27127(2) |

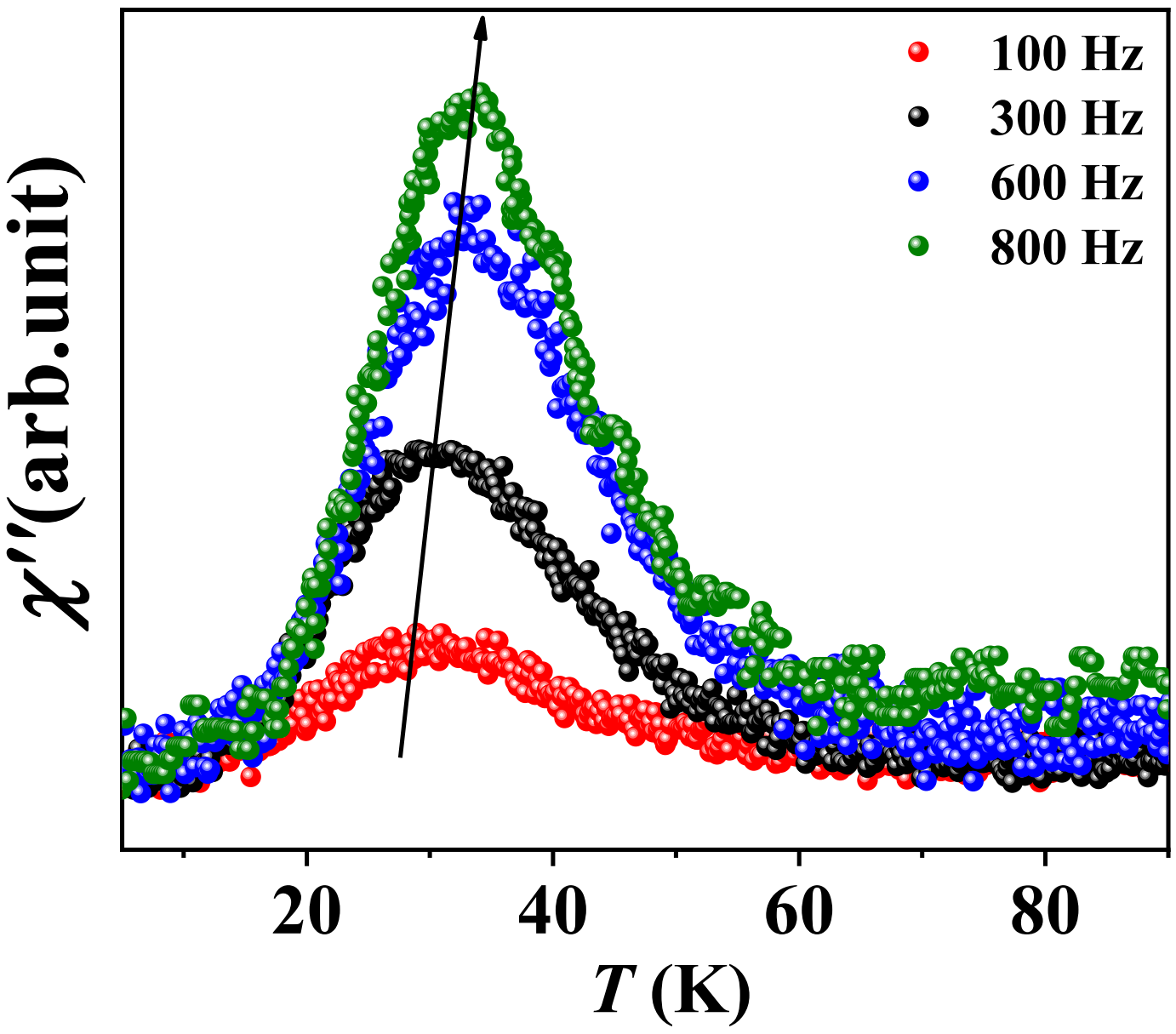

**Fig. S1**. χ'' (*T*) (imaginary part of ac susceptibility) curves at different frequencies.

χ'' (*T*) curves exhibit a broad peak, which is frequency-dependent. The peak maximum shifts to higher temperatures, and its amplitude increases with increasing frequency, which is a typical feature of the onset of a glassy state. This is in contrast to long-range magnetic order, where no frequency dispersion is observed. Thus, a re-entrant glassy state is observed at temperatures lower than $T_N$ in this system.

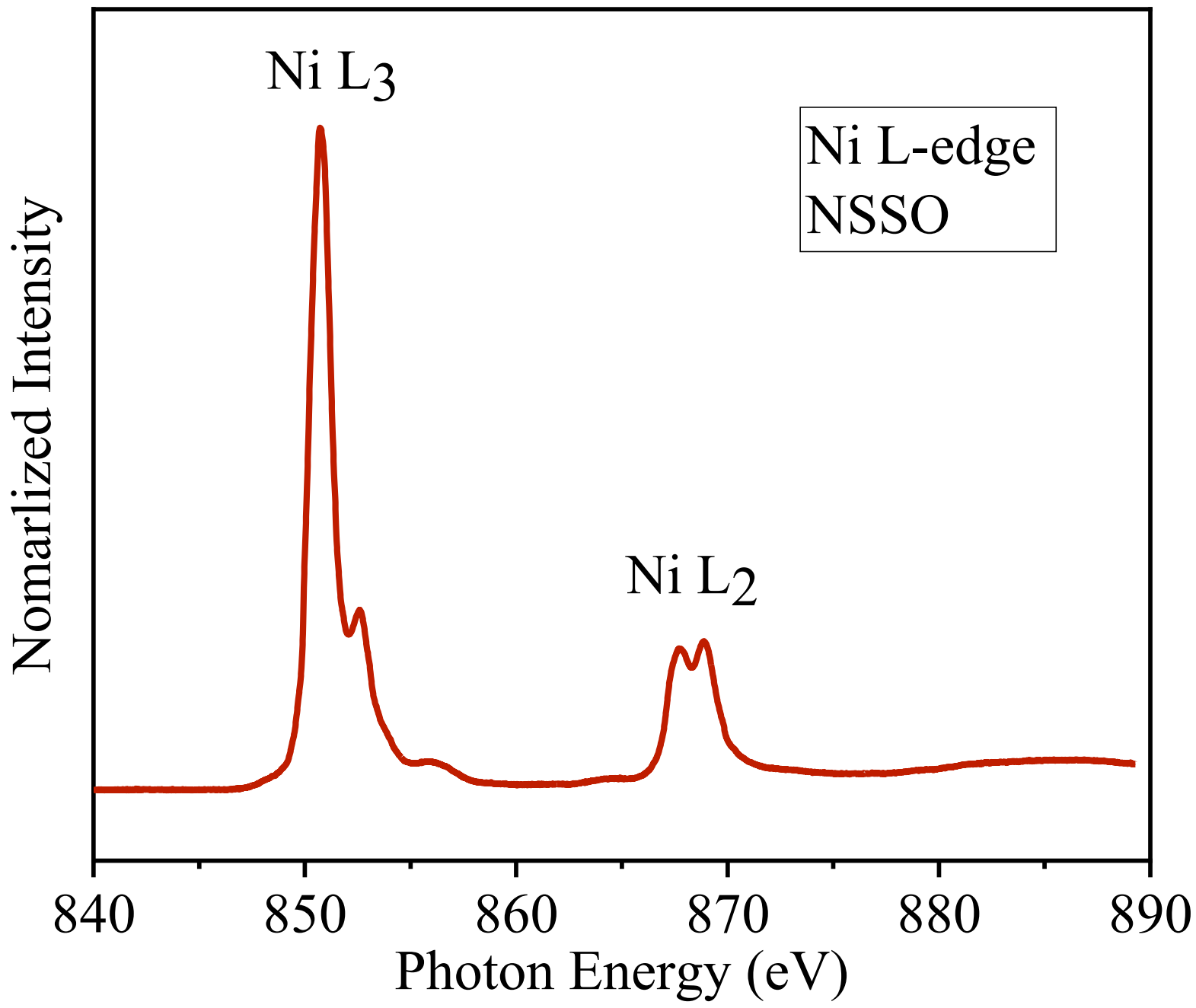

**Fig. S2.** The X-ray absorption spectrum (XAS) for Ni 2*p* at the $L_2$, $L_3$ edges, recorded at 300K. The line shape, spin-orbit coupling splitting energy, and the peak positions of the spectrum clearly suggest for the $Ni^{2+}$ oxidation state [1].

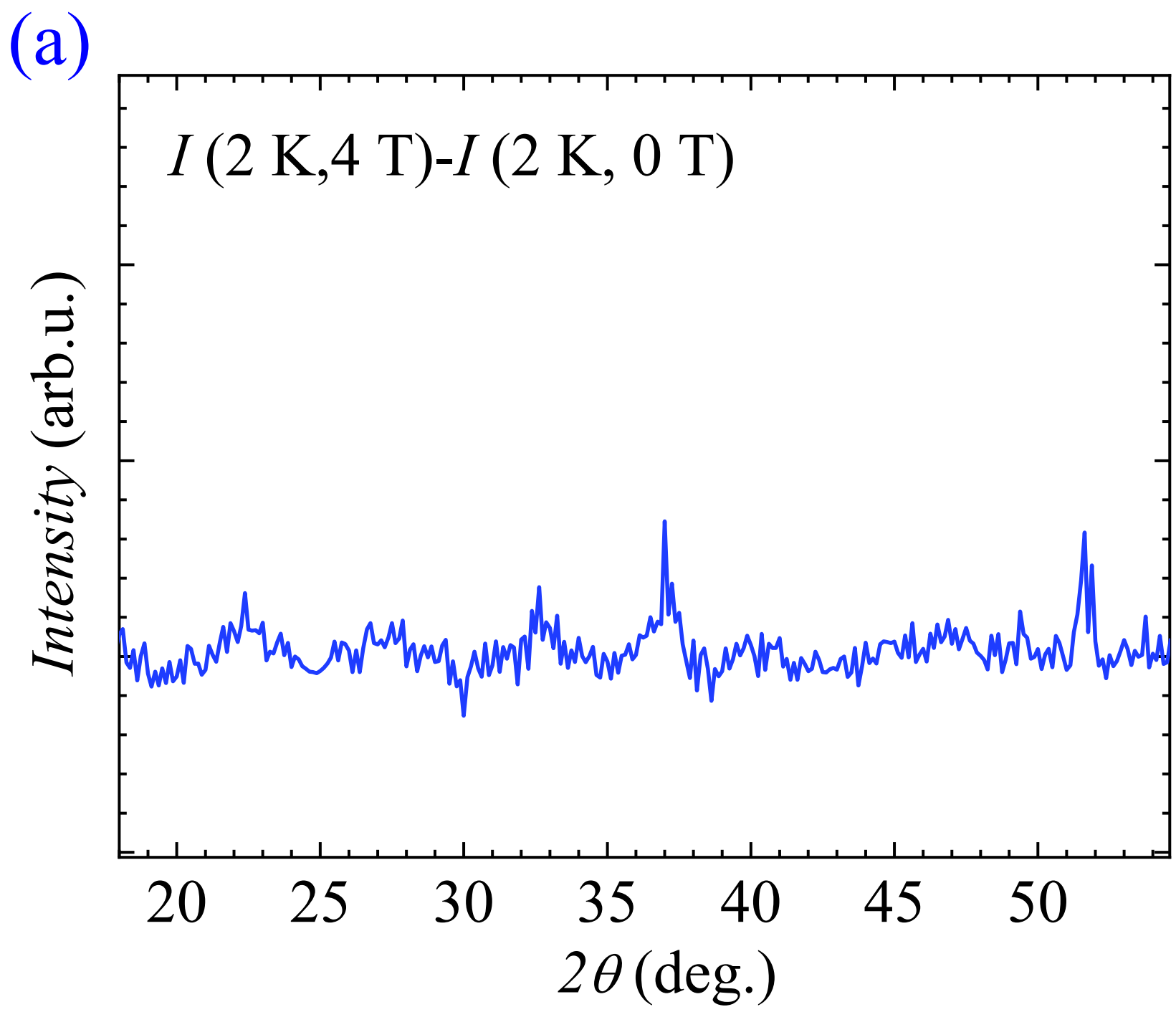

**Fig. S3** (a) Difference in 2 K NPD data measured in zero-field compared to 4 T. The diffraction profiles show no significant changes, indicating a stable spin structure under magnetic fields up to 4 T.

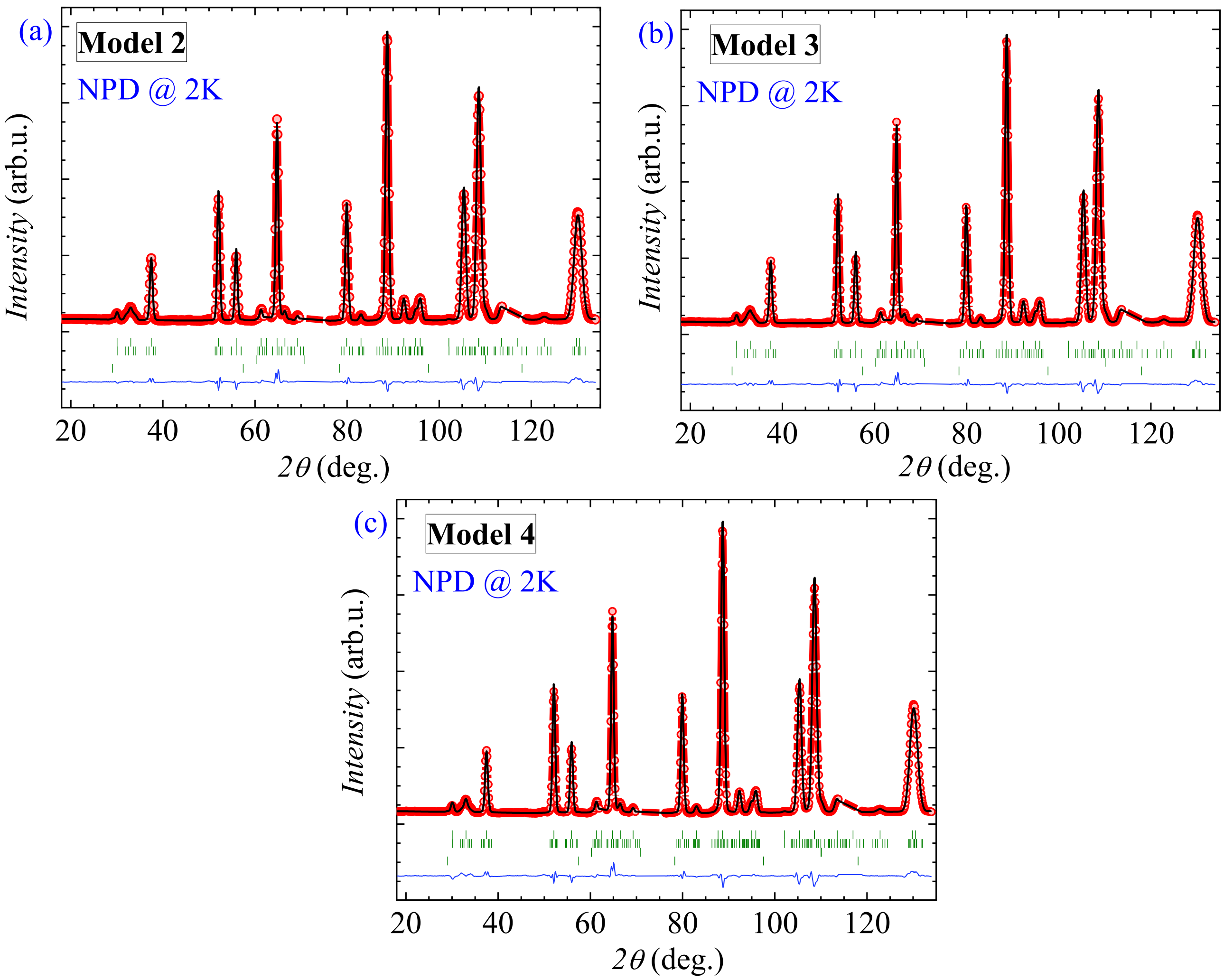

**Fig. S4.** (a)-(c) Fits (black line) to NPD pattern of $Ni_2ScSbO_6$ collected at 2 K (red data points) using different models of spin orderings. The blue lines represent the difference profiles. The four top-to-bottom green vertical bars indicate the Bragg reflection positions for $Ni_2ScSbO_6$ (nuclear and magnetic) and NiO (nuclear and magnetic) impurity, respectively.

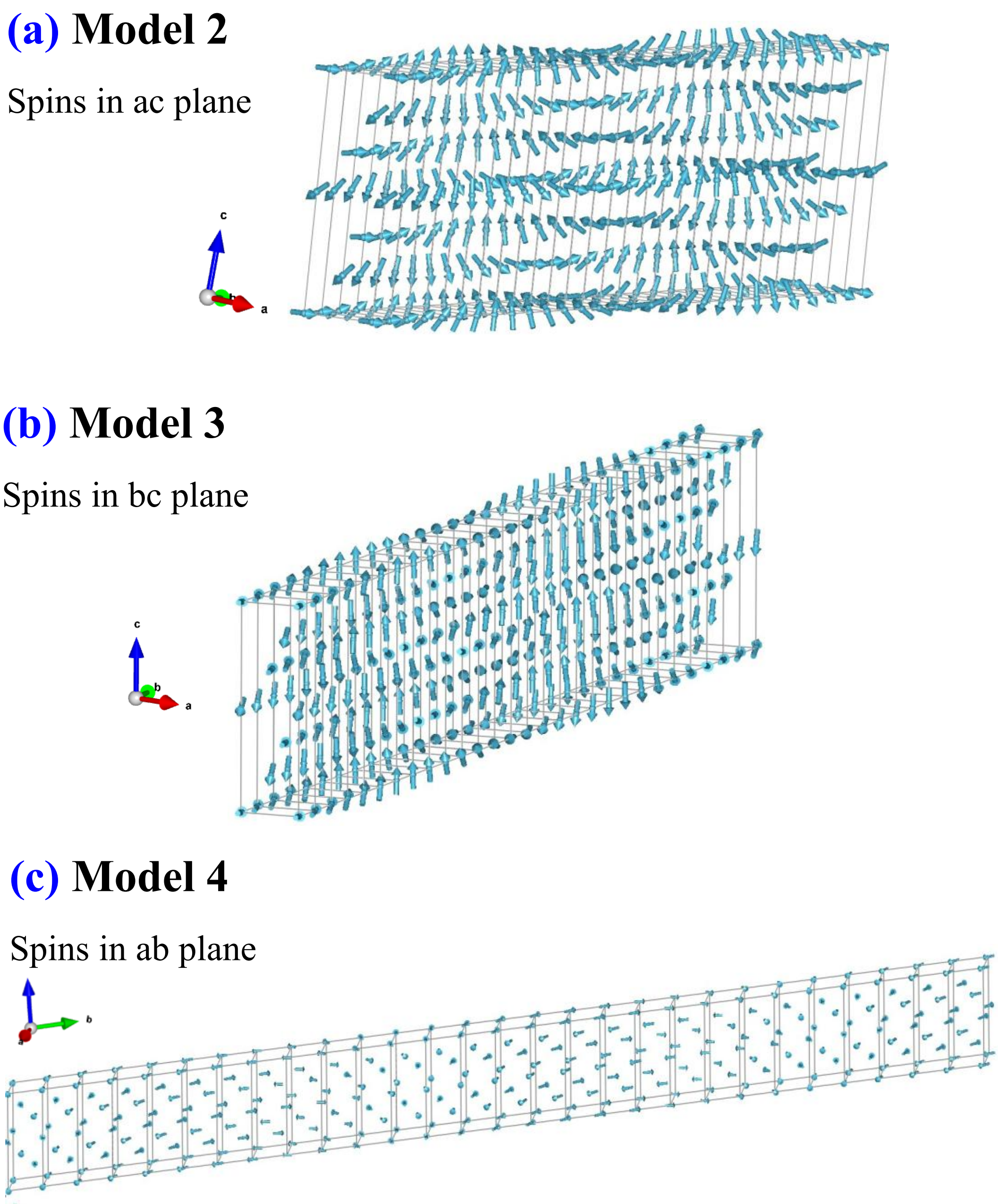

**Fig. S5.** (a)-(c) Schematic representations of the spin structures derived from various model fittings of the NPD data for Ni2ScSbO6 collected at 2 K.

**Table SII**: Comparison of the Raman modes reported for $Ni_3TeO_6$ and $Ni_2InSbO_6$ with those observed for $Ni_2ScSbO_6$.

| $Ni_3TeO_6$ Calculated mode frequency Ref: [2] | $Ni_3TeO_6$ Observed Raman mode frequencies at 300 K (in $cm^{-1}$) Ref: [2] | Symmetry | $Ni_2InSbO_6$ (TO) (in $cm^{-1}$) Ref: [3] | $Ni_2InSbO_6$ (LO) (in $cm^{-1}$) Ref: [3] | $Ni_2ScSbO_6$ (This work) |
|---|---|---|---|---|---|
| 167 | 155.74 | *A* | 155 | 158 | 164 M, 184 P1 (F) |
| 205 | 206.155 | *A* | 204 | 210 | 223 P2 (F) |
| 209 | | *E* | 192 | | 238 P3 (W) |
| 217 | | *E* | | | 250 P4 (W) |
| 264 | | *A* | | | |
| 265 | 269.60 | *E* | 271 | | 290 P5 (W) |
| 297 | 299.97 | *E* | 302 | | 312 P6 (S) |
| 343 | 340.58 | *E* | 351 | | 330 P7 (F) |
| 353 | 354 | *A* | 342 | 389 | 360 P8(W), 389 P9 (W) |
| 399 | | *A* | | 419 | 415 P10 (S) |
| 425 | 419.9229 | *E* | 423 | | 427 P11 (W) |
| 426 | 435.134 | *A* | 444 | | 445 P12 (W) |
| 488 | 506 | *E* | 484 | | 470 P13 (W), 488 P14(W) |
| 512 | 539.466 | *A* | 539 | 563 | 548 P15 (W), |
| 562 | 546.437 | *E* | | | 552 P16 (W), 570 P17 (S) |
| 601 | 632.51 | *A* | 621 | 662 | 628 P18 (F), 651 P19 (W) |
| 622 | 668.87 | *E* | 673 (A) | 696 (A) | 687 P20 (S) |
| 653 | 699.471 | *A* | 702 | 706 | 723 P21 (Max), 736 P22 (W) |

*Key: M-Magnon mode; Max-Maximum; W-Weak; S-Strong; F-Faint.

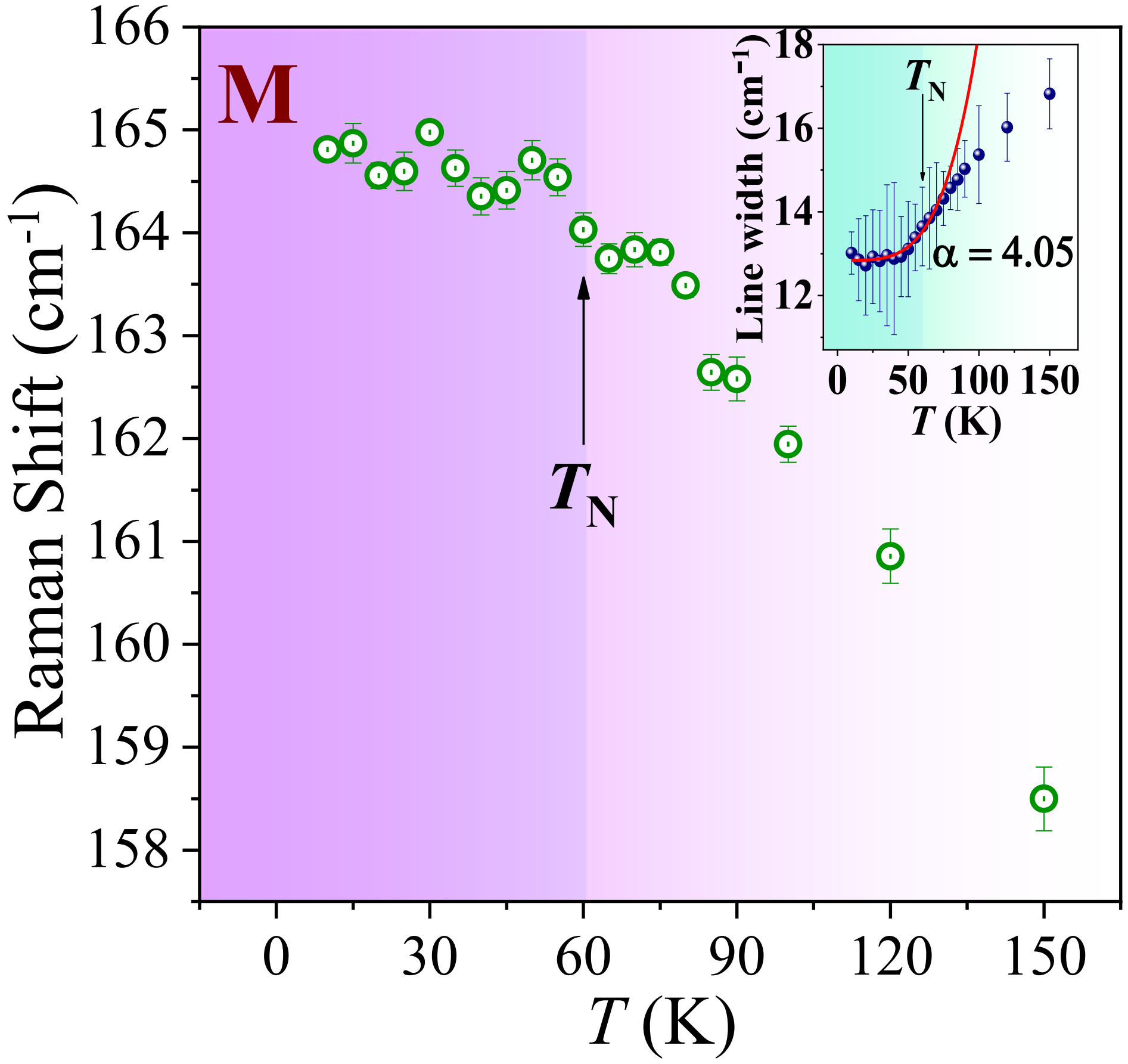

**Fig. S6**. Temperature variation of the *M* mode (possible magnon mode) frequency. Inset: corresponding thermal variation of the linewidth of the *M* mode. The red curve represents a fit using a $T^{\alpha}$ law.

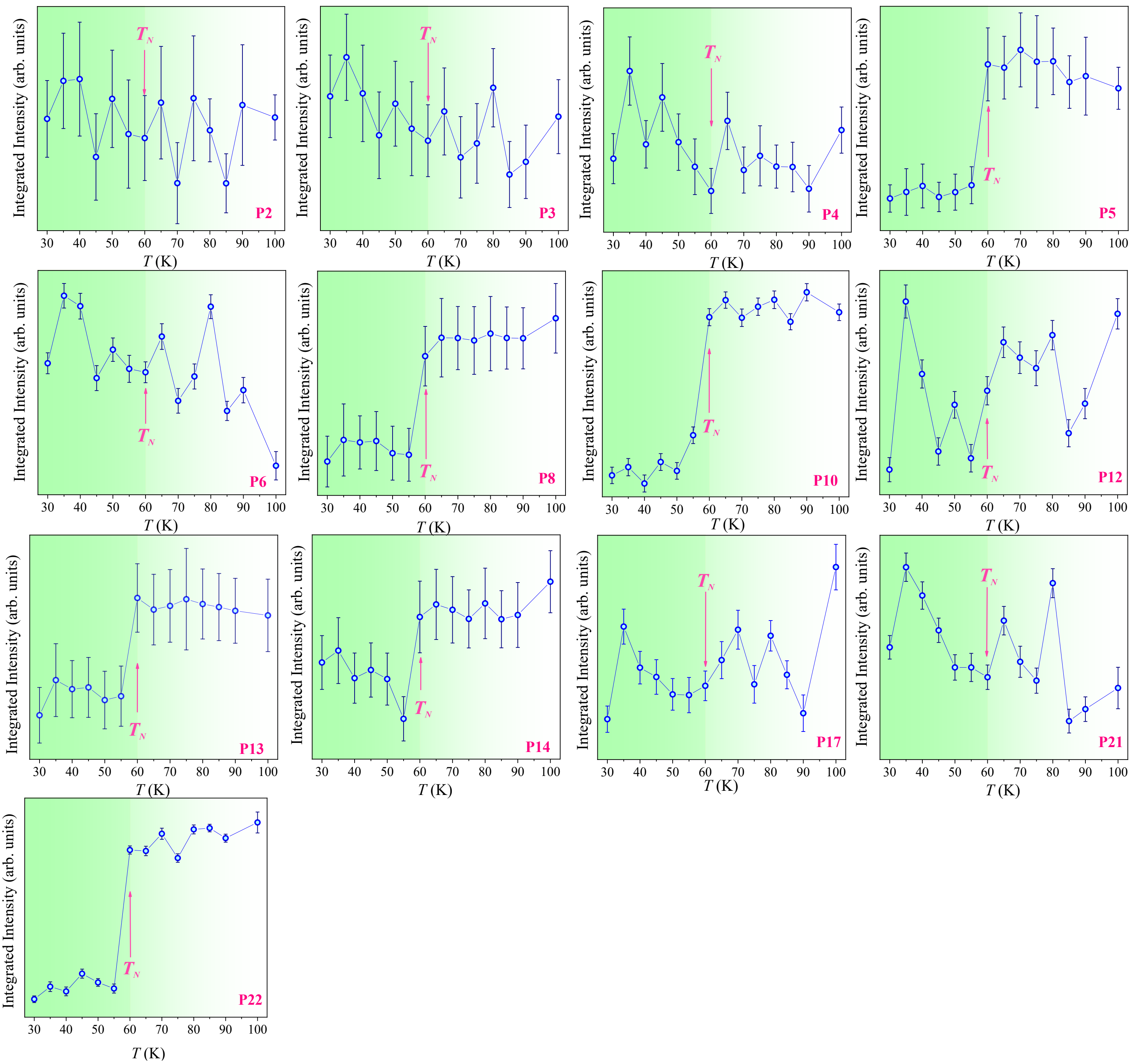

**Fig. S7.** Temperature dependence of integrated intensities of selected Raman modes.

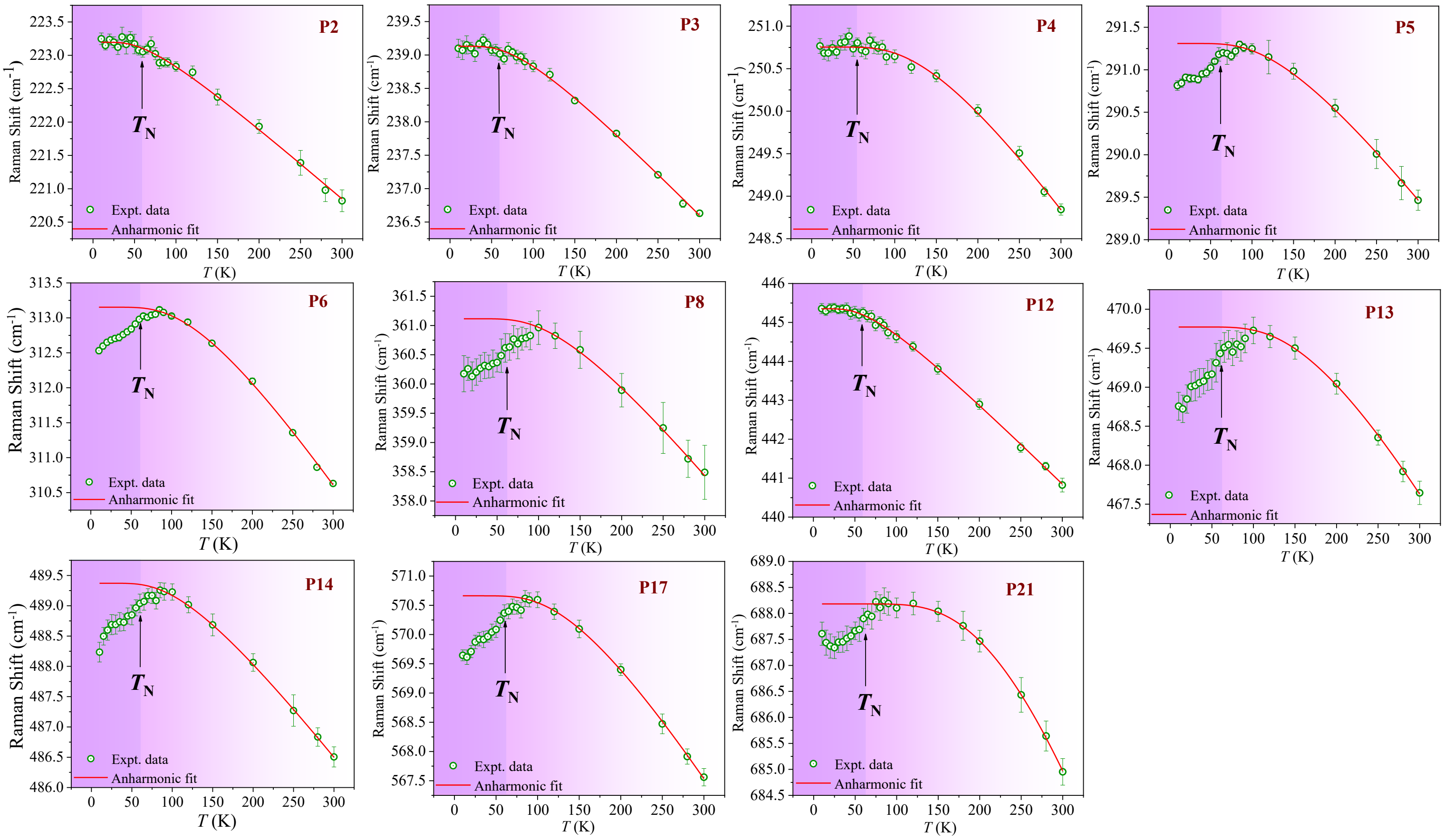

**Fig. S8.** Temperature dependence of frequencies of selected Raman modes.

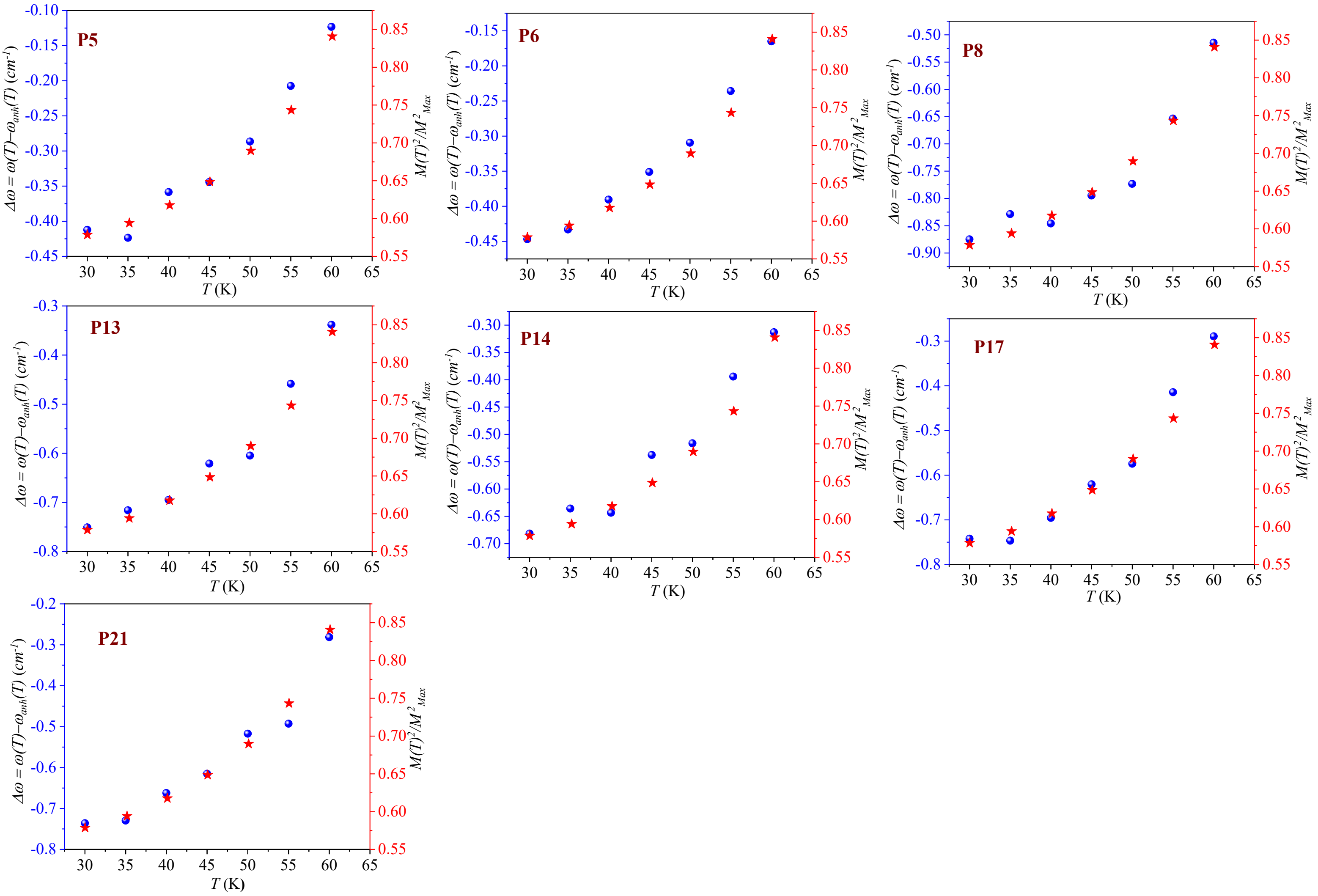

**Fig. S9.** Temperature dependence of $\Delta\omega$ and $\{M(T)/M_{Max}\}^2$ for selected Raman modes.

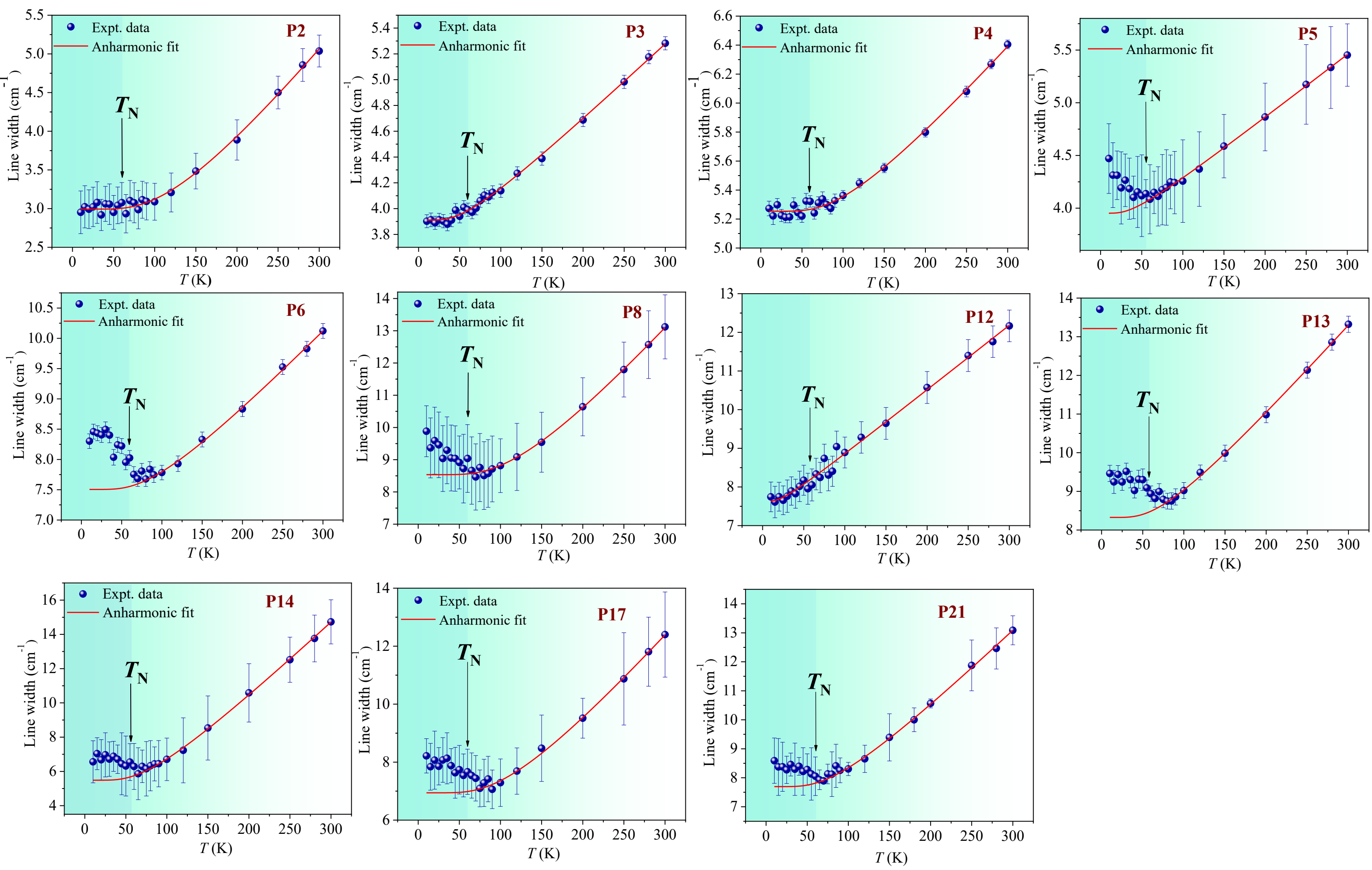

**Fig. S10.** Temperature dependence of line-widths of selected Raman modes.

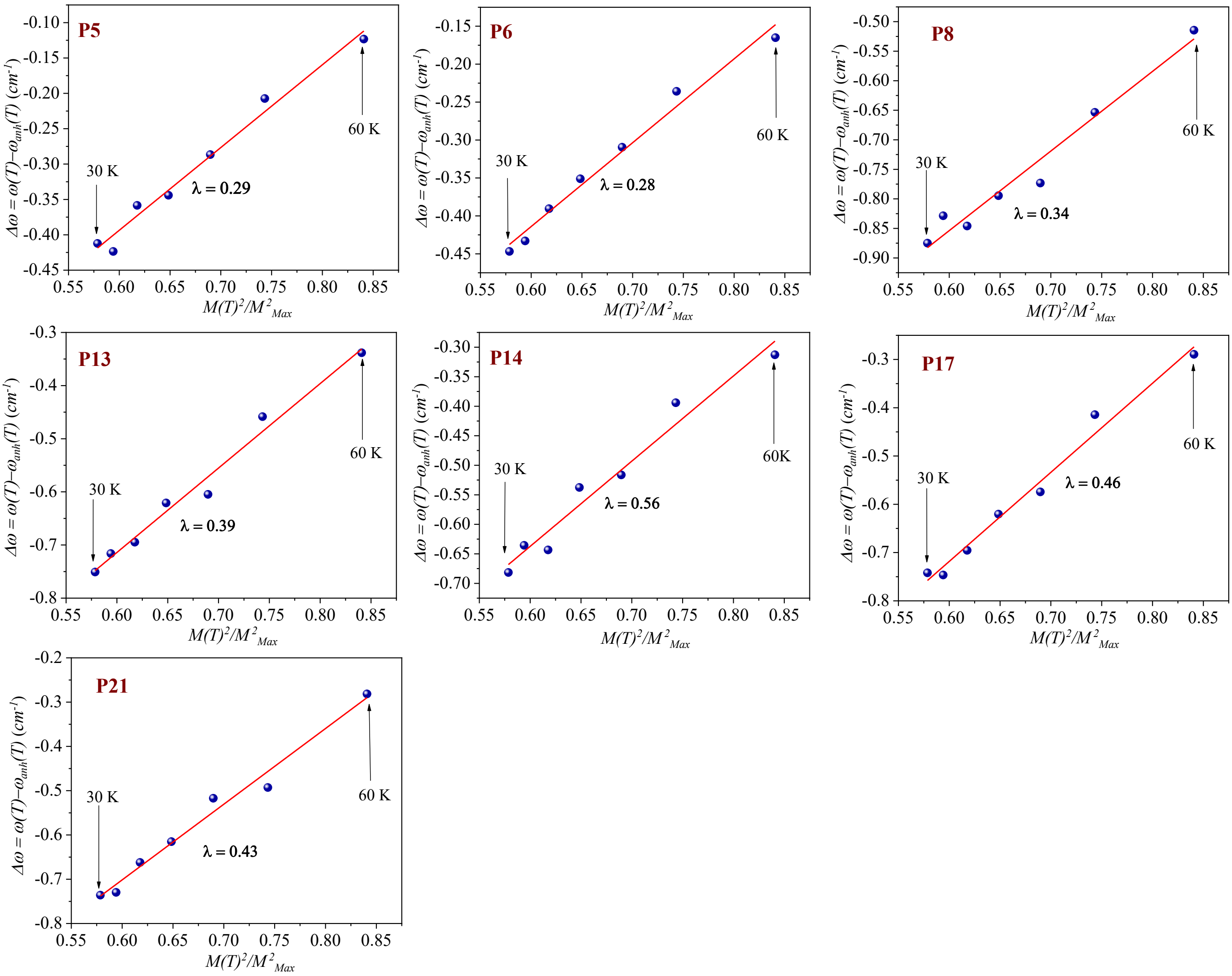

**Fig. S11.** $\Delta\omega$ vs. $\{M(T)\,/\,M_{Max}\}^2$ curves and their linear fits used to determine the SPC constant λ for various modes.

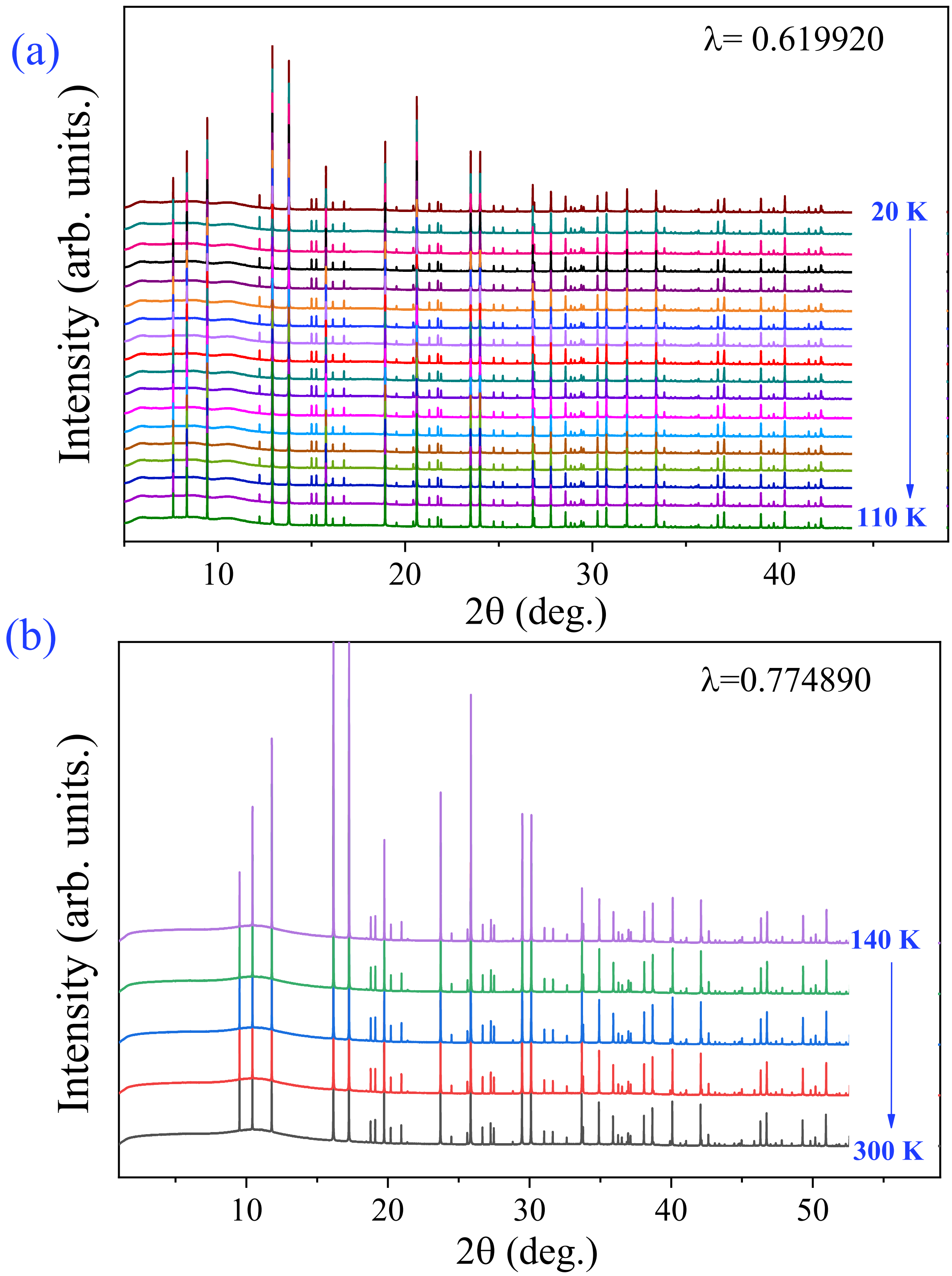

**Fig. S12.** SXRD patterns (a) collected at temperatures ranging from 20 K to 110 K for $\lambda_i = 0.619920$ and (b) from 140 K to 300 K for $\lambda_i = 0.774890$.

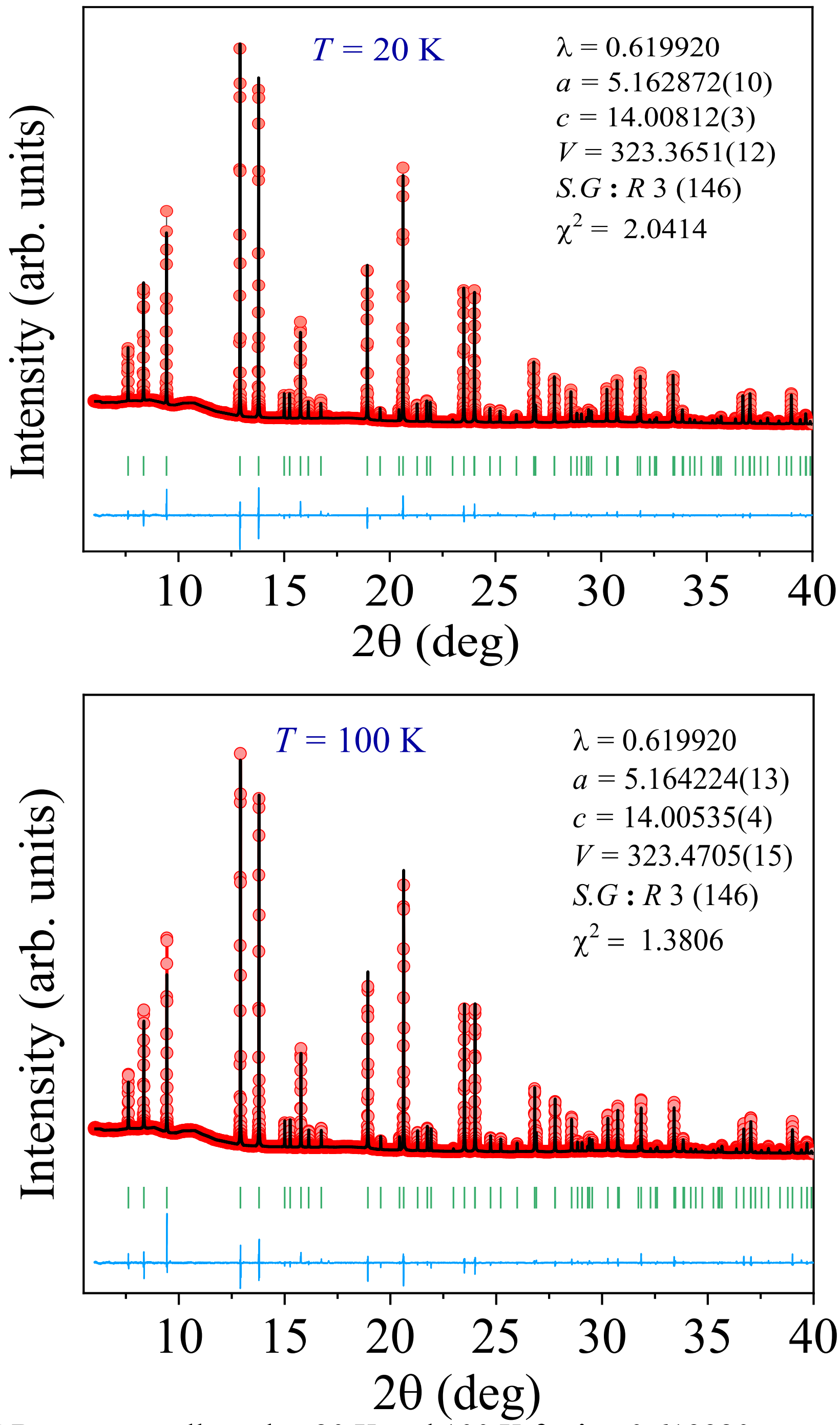

**Fig. S13.** Fitted SXRD patterns collected at 20 K and 100 K for λ = 0.619920.

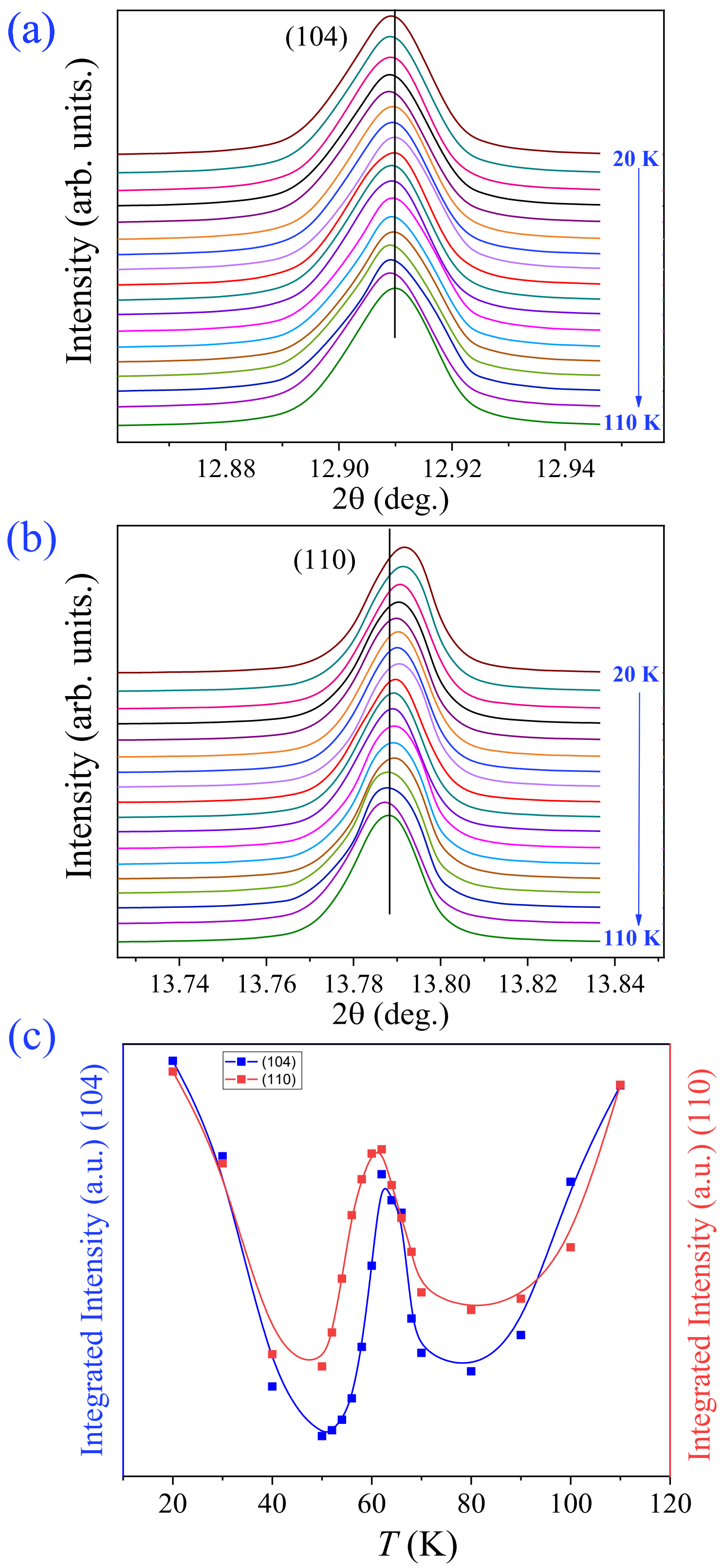

**Fig. S14.** (a) and (b) Evolution of 104 and 110 SXRD peaks in the temperature range around $T_N$. (c) Integrated intensity of 104 and 110 SXRD peaks as a function of $T$.

**Table SIII** Fitting parameters obtained from the thermal expansion fit to the $T$ dependence of the lattice parameters.

| Lattice Parameters | Temperature range (K) | |
|---|---|---|
| ***a* = *b* (Å)** | 300 K – 60 K | $a_0$ = 5.16406(73) Å<br>$B$ = 1.98(6)<br>and $d$ = −685(74) |
| ***c* (Å)** | 300 K – 60 K | $c_0$ = 14.00545(68) Å<br>$f$ = 5.7(1)<br>and $g$ = 1040 (51) |
| ***V* (Å$^3$)** | 300 K – 60 K | $V_0$ = 323.4285(59) Å$^3$<br>$A$ = 0.0123(13)<br>and $\theta_D$ = 542(25) K |

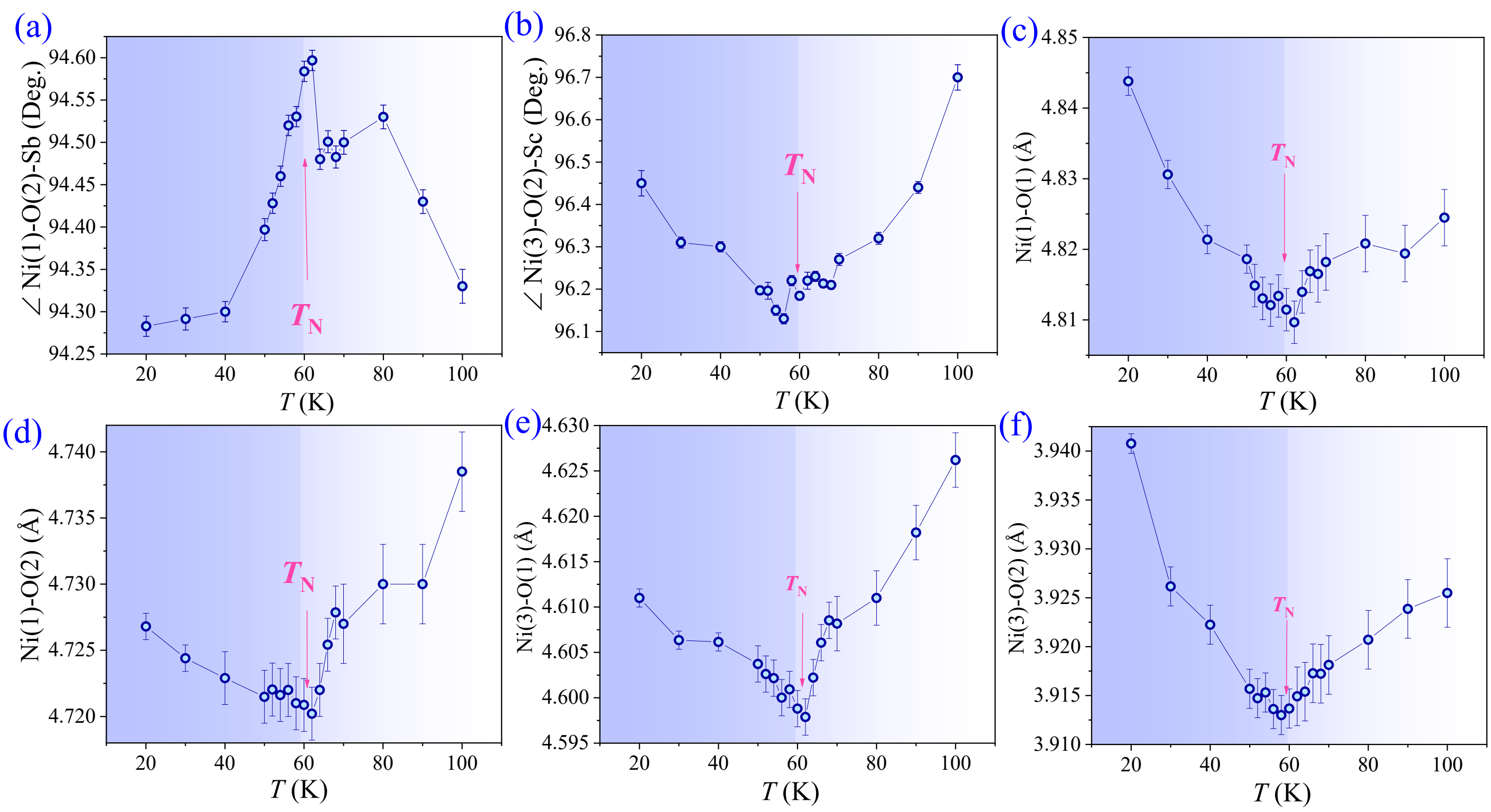

**Fig. S15.** Temperature dependence of selected bond angles and bond lengths, showing anomalous behavior around $T_N$.

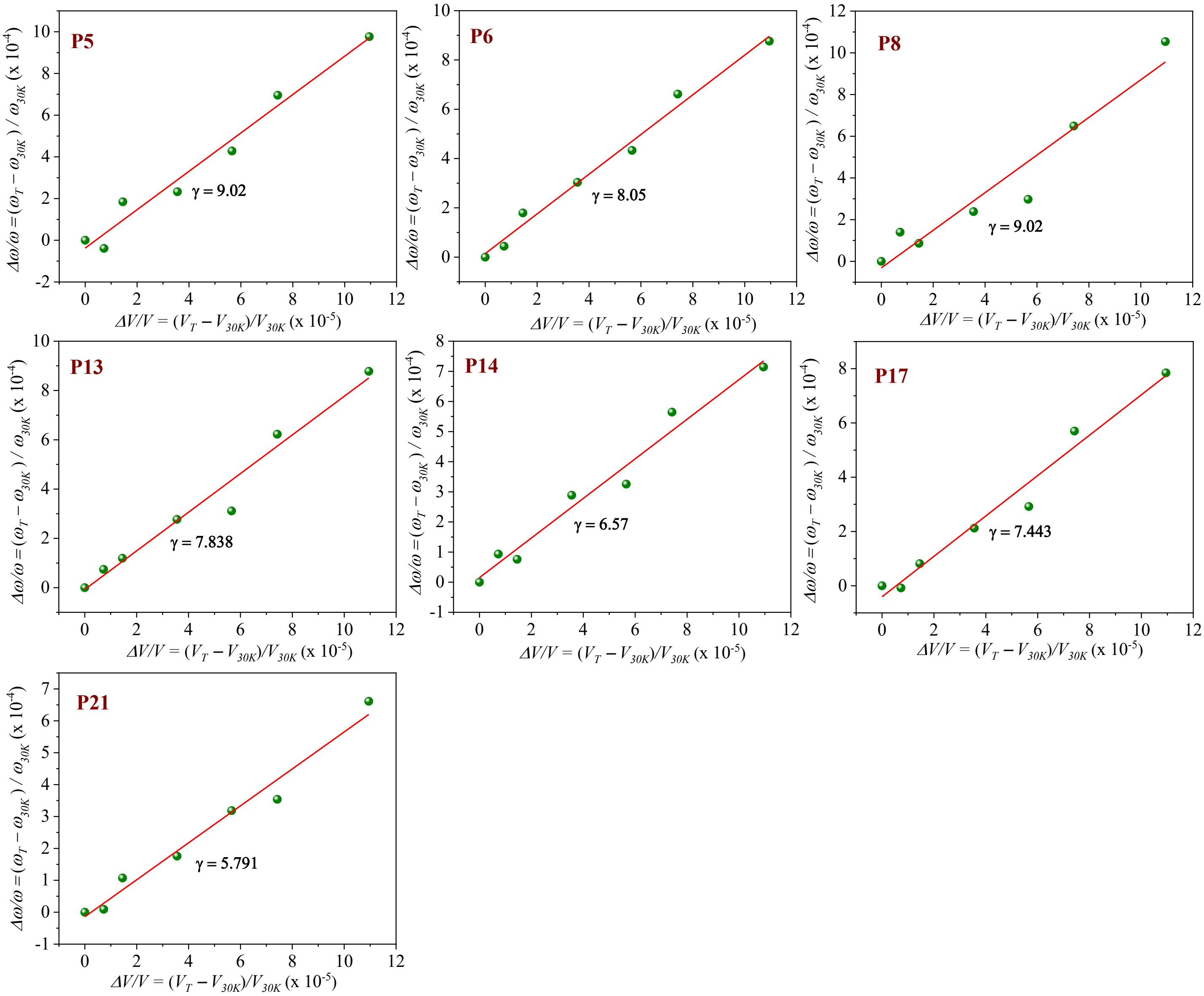

**Fig. S16.** (Δω/ω) vs. (ΔV/V) curves for various phonon modes. The Grüneisen parameters, γ, are derived from linear fits (red).